\documentclass[fleqn,usenatbib]{mnras}

\usepackage{newtxtext,newtxmath}

\usepackage[T1]{fontenc}

\DeclareRobustCommand{\VAN}[3]{#2}
\let\VANthebibliography\thebibliography
\def\thebibliography{\DeclareRobustCommand{\VAN}[3]{##3}\VANthebibliography}

\usepackage{graphicx}	
\usepackage{amsmath}	
\usepackage{hyperref}
\usepackage{subcaption}
\usepackage{orcidlink}

\newcommand{\aref}[1]{\hyperref[#1]{Appendix~\ref{#1}}}

\usepackage{afterpage}

\usepackage[normalem]{ulem}

\newcommand{\hi}{\ifmmode{\rm HI}\else{H\/{\sc i}}\fi}
\newcommand{\htwo}{H\textsubscript{2}}

\newcommand{\nhi}{\ensuremath{N_\hi}}

\newcommand{\shi}{\ensuremath{\Sigma_\mathrm{at}}}
\newcommand{\shii}{\ensuremath{\Sigma_\mathrm{mol}}}

\newcommand{\kms}{\ensuremath{\mathrm{km\, s}^{-1}}}
\newcommand{\mpc}{\ensuremath{\mathrm{M}_\odot\: \mathrm{pc}^{-2}}}

\title[Atomic-to-molecular transition in the MW wind]{Direct observations of the atomic-molecular phase transition in the Milky Way's nuclear wind}

\author[Noon et al.]{Karlie A. Noon$^{\orcidlink{0000-0002-9699-6863}}$,$^1$\thanks{E-mail: karlie.noon@anu.edu.au}
Mark R. Krumholz$^{\orcidlink{0000-0003-3893-854X}}$,$^{1,2}$
Enrico M. Di Teodoro$^{\orcidlink{0000-0003-4019-0673}}$,$^{3}$
Naomi M. McClure-Griffiths$^{\orcidlink{0000-0003-2730-957X}}$,$^{1}$
\newauthor Felix J. Lockman$^{\orcidlink{0000-0002-6050-2008}}$$^{4}$
and Lucia Armillotta$^{\orcidlink{0000-0002-5708-1927}}$$^{5}$
\\
$^{1}$Research School of Astronomy and Astrophysics, The Australian National University, Canberra, Australian Capital Territory, Australia.\\
$^{2}$ARC Centre of Excellence for Astronomy in Three Dimensions (ASTRO3D), Canberra, ACT~2611, Australia\\
$^{3}$Dipartimento di Fisica e Astronomia, Università degli Studi di Firenze, via G. Sansone 1, 50019 Sesto Fiorentino, Firenze, Italy\\
$^{4}$Green Bank Observatory, Green Bank, WV 24944, USA\\
$^{5}$Department of Astrophysical Sciences, Princeton University, Princeton, NJ 08544, USA
}

\date{Accepted XXX. Received YYY; in original form ZZZ}

\pubyear{2023}

\begin{document}
\label{firstpage}
\pagerange{\pageref{firstpage}--\pageref{lastpage}}
\maketitle

\begin{abstract}
Hundreds of high-velocity atomic gas clouds exist above and below the Galactic Centre, with some containing a molecular component. However, the origin of these clouds in the Milky Way's wind is unclear. This paper presents new high-resolution MeerKAT observations of three atomic gas clouds and studies the relationship between the atomic and molecular phases at $\sim 1$ pc scales. The clouds' atomic hydrogen column densities, $N_{\mathrm{HI}}$, are less than a $\mbox{few}\times 10^{20}$ cm$^{-2}$, but the two clouds closest to the Galactic Centre nonetheless have detectable CO emission. This implies the presence of \htwo{} at levels of $N_{\mathrm{HI}}$ at least a factor of ten lower than in the typical Galactic interstellar medium. For the cloud closest to the Galactic Centre, detectable CO coexists across the entire range of H\textsc{i} column densities. In contrast, for the intermediate cloud, detectable CO is heavily biased toward the highest values of $N_{\mathrm{HI}}$. The cloud most distant from the Galactic Centre has no detectable CO at similar $N_{\mathrm{HI}}$ values. Moreover, we find that the two clouds with detectable CO are too molecule-rich to be in chemical equilibrium, given the depths of their atomic shielding layers, which suggests a scenario whereby these clouds consist of pre-existing molecular gas from the disc that the Galactic wind has swept up, and that is dissociating into atomic hydrogen as it flows away from the Galaxy. We estimate that entrained molecular material of this type has a $\sim \mathrm{few}-10$ Myr lifetime before photodissociating.
\end{abstract}




\section{Introduction}

\label{sec:intro}
A large-scale multiphase bipolar outflow emanates from the Milky Way's (MW) Galactic Centre (GC), driven either by past nuclear activity or by past periods of intense star formation taking place in the Central Molecular Zone \citep[e.g.][]{BlandHawthorn2003, Su2010, Crocker2015, Yang2022}. This nuclear wind includes components visible at all observable wavelengths ranging from the very hot X-ray emitting phase at $T \sim 10^{6-8}$ K \citep[e.g.][]{Koyama1989, Snowden1997, Su2010}, to the radio-bright cold atomic and molecular gas phase at just $\sim 10-100$ K \citep[e.g.][]{Sofue1984, Lockman1984, McClure-Griffiths2013, DiTeodoro2020, Cashman2021}. Gamma-ray observations by \cite{Su2010}, \citet{Dobler2010}, and \citet{Carretti2013} reveal that the wind also includes a relativistic component, called the Fermi Bubbles, that extends up to 10 kpc above and below the GC. The various components of the wind carry energy, baryons, and metals from the interstellar medium (ISM) to the Galaxy's circumgalactic medium (CGM).

The MW's wind is interesting in its own right, but it also plays an important role in the study of galactic winds (GWs) more broadly. Since it is at a distance of only $\sim$ 8 kpc, it serves as the closest laboratory to study the physics of the GWs responsible for carrying large amounts of star-forming material out of galactic discs, regulating star formation and significantly impacting galactic evolution. The mechanisms that drive the winds, their physical properties, the spatial and mass distribution of their multiphase components, and their impacts on the host galaxy are still poorly understood \citep{Veilleux2020}. 

One problem for which the MW wind can provide special insight is the origin of cold atomic and molecular gas in outflows. Such gas has been seen in numerous extra-galactic observations \citep[e.g.][]{Morganti1998, Seaquist2001, Greve2004, Leroy2015, Martini2018}, and conceivably dominates the total wind mass flux \citep[e.g.,][]{Yuan2023}, but its presence is difficult to understand since simulations suggest that the material driving hot winds should rapidly shock-heat cold gas to high temperatures \citep[e.g.,][]{Schneider2017}. A large number of models have been proposed to explain the survival of cool gas \citep[e.g.,][]{McCourt2015, Schneider2020, Huang2020, Kanjilal2021}, but these have proven difficult to test in observations due to the limited spatial resolution available in extragalactic data, which precludes resolving individual clouds entrained in the wind. By contrast, modern radio telescopes can resolve $\sim 1$ pc spatial scales in the MW wind, making it possible to study the bulk properties and morphology of individual cold clouds suspended in it.

A second, closely related problem where the MW again offers unique advantages is in understanding how the phase structure of winds evolves with distance from the host galaxy. Large amounts of atomic and molecular gas are entrained in galactic outflows, yet galaxy circumgalactic media are mostly comprised of ionised gas \citep{Werk2014}. This implies that a phase transition must occur, and there are claims that this transition has been observed in the nearby starburst galaxy M82 \citep{Leroy2015, Martini2018}. However, the interpretation of the data as showing evidence for a phase transition remains controversial due to uncertainties about how the velocities of different chemical components of the wind evolve with height \citep{Yuan2023}. Similarly, there are theoretical model predictions for how galactic wind phases vary with galactocentric distance \citep[e.g.,][]{Fielding2022}, but these have not yet been subjected to strong observational tests. Resolved observations of individual clouds at a range of galactocentric distances would be ideal for this purpose, but at present they are not available in external galaxies. 

By contrast, a number of clouds suitable for studies of this type are known in the MW. \cite{Lockman1984} and \cite{Lockman2016} noted a large void of neutral hydrogen gas above and below the disc in the inner 2 kpc of the MW, potentially indicative of large-scale winds sweeping the region free of cold gas \citep{Bregman1980}. Within this void region, \cite{McClure-Griffiths2013} observed 86 anomalous atomic hydrogen (\hi) clouds above and below the GC with velocity distributions consistent with being entrained in an outflowing wind. Since then, \cite{DiTeodoro2018} and \cite{Lockman2020}, combined with \cite{McClure-Griffiths2013}, have detected a total population of approximately 200 cold, dense \hi\ clouds that have kinematics consistent with being entrained in a nuclear outflow from the GC. From this population, \cite{DiTeodoro2020} observed that two clouds were associated with small clumps of cold molecular matter. This represents the first observations of molecular matter entrained in the MW's nuclear wind. 

Three plausible scenarios have been proposed to explain the origin and survival of these molecular structures, each carrying with it different assumptions about where cold gas comes from and how gas phases change as the wind flows. One scenario is that the nuclear wind has blown chunks of molecular clouds from the disc that, as they flow outward, dissociate \citep[e.g.][]{Espada2010, Saito2022, Saito2022b} into \hi~(and possibly thereafter ionise) due to the surrounding radiation and gas. In a second scenario, the hot wind has entrained atomic gas from the disc and compressed it into molecular form due to the high pressure. Finally, a third scenario is that the wind was originally composed mostly of hot, ionised material that has subsequently condensed and cooled into atoms and molecules due to thermal instabilities \citep[e.g.][]{Schneider2018, Thompson2020, MacCagni2021}. In the first of these scenarios, there is also a possibility that the wind has entrained atomic hydrogen along with the molecular matter, as atomic and molecular gas have long been observed to be correlated \citep[e.g.][]{Simonson1973, Wannier1983, Andersson1992}, and on theoretical grounds, all molecular clouds in the disc are expected to be protected by a dusty atomic envelope \citep{Krumholz2008, Krumholz2009, McKee2010}. 
Resolved observations of individual clouds can differentiate between these scenarios, which in turn will provide useful insight into how wind phase structures evolve as gas is blown away from the disc into the circumgalactic medium. The insight we gain from the MW can help us understand winds more generally.

To further investigate the phase structure of material entrained in the MW's wind, this paper will focus on three bright \hi{} clouds identified in prior surveys \citep{McClure-Griffiths2013, DiTeodoro2018}. Two of these clouds have been observed in the $^{12}$CO(2$\rightarrow$1) emission line \citep{DiTeodoro2020}, while a third was observed but not detected, yielding informative upper limits. Low$-J$ CO lines are an accessible and widely used proxy for molecular gas, allowing us to explore the chemical states of the clouds. We complement these data with new, high-resolution \hi\ interferometric observations from MeerKAT, allowing us for the first time to study the phase structure of MW wind clouds with high, matched resolution for the atomic and molecular phases. 

The remainder of this paper is structured as follows: first, in \autoref{sec:obs} we present both the new MeerKAT 21 cm observations and CO data, and describe how we extract column densities and masses for the atomic and molecular components. \autoref{sec:analysis} compares the \hi{} and CO data, both from a morphological and quantitative point of view, and analyses the chemical state of the gas. We interpret these observations in light of the scenarios described above in \autoref{sec:disc}, and summarise our findings in \autoref{sec:conc}. 

\section{Observations and data reduction}
\label{sec:obs}

In this section, we outline the observations of the three target clouds that are the focus of this paper. Hereinafter, we will refer to these sources as C1, C2, and C3. We describe the \hi{} observations and processing in \autoref{sec:hiobs} and the corresponding procedures for the CO data in \autoref{sec:coobs}. We summarise some features of the observations in \autoref{tab:sourceinfo}.

\begin{table*}
	\centering
	\caption{Measured and derived properties of observed clouds. Galactic Longitudes $l$ and latitudes $b$ record the centres of the fields of view. V$_\mathrm{LSR}$ is the \hi{} central local standard of rest velocity, $N_{\mathrm{H,\hi{},peak}}$ and $N_{\mathrm{H,H_2,peak}}$ are the peak derived column density of the \hi{} and \htwo{} data given in units H nuclei per cm$^2$ -- thus the quantity we report for the H$_2$ data is the number of H \textit{nuclei} per unit area we infer, not the number of H$_2$ molecules, so the H~\textsc{i} and H$_2$ numbers are directly comparable. $M_{\mathrm{at}}$ is the derived \hi{} mass from \autoref{eqn:masshi}, and  $M_{\mathrm{mol}}$ is the corresponding derived H$_2$ mass from \autoref{eqn:massh2}.}
 \label{tab:sourceinfo}
	\begin{tabular}{lccccccr}
		\hline
		Cloud ID&\emph{l}&\emph{b}&V$_\mathrm{LSR}$&$N_{\mathrm{H,\hi{},peak}}$&$N_{\mathrm{H,H_2,peak}}$&$M_\mathrm{at}$&$M_\mathrm{mol}$\\ 
                & ($^\circ$) & ($^\circ$) & (\kms{}) & (cm$^{-2}$) & (cm$^{-2}$) & (M$_\odot$) & (M$_\odot$)\\
		\hline
		C1 & $-1.85$ & $-3.87$ & 165 & $1.8\times 10^{20}$ & $1.38\times 10^{21}$  & 342 & 394 \\
		C2 & $-2.43$ & 5.55 & 264 & $2.3\times 10^{20}$ & $1.00\times 10^{21}$ & 721 & 571 \\
            C3 & 13.25 & 7.0 & $-$99 & $1.5\times 10^{20}$ & - & 3429 & - \\
		\hline
\end{tabular}
 \end{table*}
\subsection{\hi{} Observations}
\label{sec:hiobs}
We carried out high-resolution observations of the 21 cm \hi\ emission line in December 2020 using the MeerKAT radio interferometer (see \cite{Jonas2016} for a description of the telescope). We observed the three objects over a total integration time of 3 hours using a single pointing for each cloud centred at Galactic coordinates $(l,b)=(-1.86^{\circ}, -3.84^{\circ})$, $(l,b)=(-2.42^{\circ}, 5.56^{\circ}$), and $(l,b)=(13.30^{\circ}, 7.00^{\circ})$ for C1, C2, and C3, respectively. For these observations, 59 of the 64 MeerKAT antennas were available, giving 1770 baselines. The shortest and longest baselines were 29 m and 8 km, respectively. The observations have a bandwidth of 856 MHz to 1712 MHz across 32768 channels at a spectral resolution of 26.12 kHz, corresponding to a velocity resolution of 5.5 \kms{} at the \hi\ frequency of 1420.405 MHz.

The MeerKAT pipeline calibrated the raw data. The pipeline includes flagging of radio frequency interference (RFI), bandpass calibration, and receiver gain fluctuations calibration. For an in-depth guide to the MeerKAT calibration pipeline, see \cite{Wang2021}. The South African Radio Astronomy Observatory's archive\footnote{\url{https://archive.sarao.ac.za/}} stores the calibrated $u-v$ dataset. To image these data, we use the Common Astronomy Software Applications \citep[CASA,][]{McMullin2007}, version 5.4.1-32. We first split the full $u-v$ dataset into three separate datasets, one for each source. Using the \emph{uvcontsub} task in CASA, we generate a model of the continuum sources by fitting emission-free channels to a line. The \emph{uvcontsub} function subtracts this model from the $u-v$ dataset. We use the \emph{tclean} task to image each of the three sources into separate datacubes. We use a linear interpolator, a standard gridder with a cell size of 5$''$, a common restoring beam, a multiscale deconvolver with scales $=[0, 6, 10, 20, 40, 80]$ pixels and natural weighting. To mask the images, we use an auto-multithresh mask with a side lobe threshold of 3, a noise threshold of 5, a low noise threshold of 1.5, and a negative threshold of 7. Cleaning ceases after reaching the interaction limit of 10,000 (as was the case for C1) or when the cleaning procedure reaches a stopping threshold which is given by the noise level of the dirty cube (as was the case for C2 and C3). 

At 1420 MHz, the restored beam is 24.9$''$ $\times$ 21.7$''$ with position angle $-2.8^{\circ}$. The primary beam FWHM is $67^\prime \times 67^\prime$. The root-mean-square (rms) noise ($\sigma_\mathrm{chn}$) is 304 mK, 275 mK  and 365 mK for C1, C2, and C3, respectively, per 5.5 \kms{} channel. To correct for the missing short-spacing baselines, we combine the interferometric data with Green Bank Telescope (GBT) single-dish observations from \cite{DiTeodoro2018} using CASA's \emph{feather} task with a single-dish flux scaling factor of 1. See \aref{App:compari} for details on our comparison between the MeerKAT and GBT data.

\begin{figure*}
	\includegraphics[width=16cm, height=16cm]{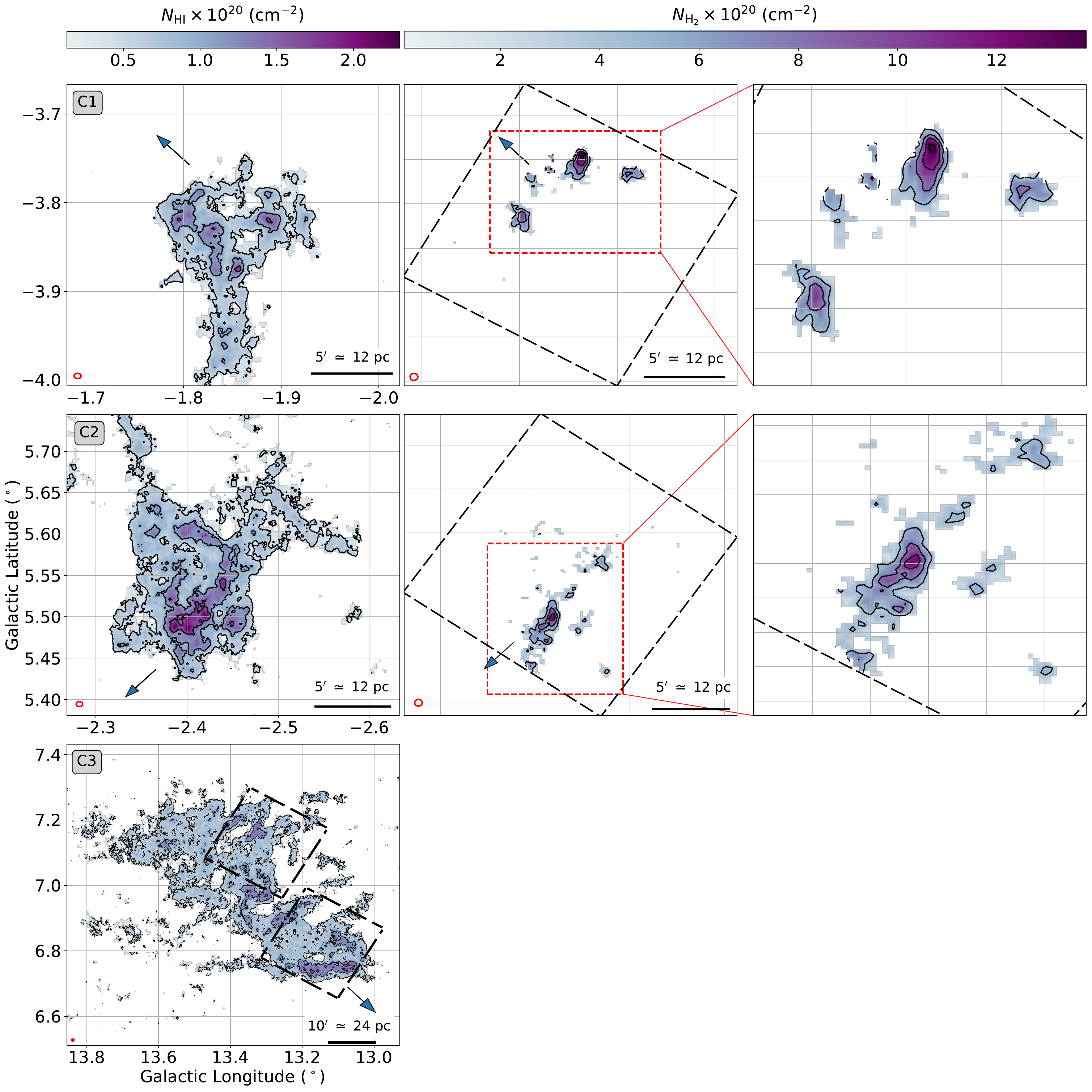}
    \caption{Column density maps derived from the MeerKAT $\sim 23^{\prime\prime}$ angular resolution \hi{} observations (left) and the APEX $\sim 28^{\prime\prime}$ $^{12}$CO(2$\rightarrow$1) observations (middle and right) following the procedure outlined in \autoref{sec:obs}. All panels show the inferred number density of H nuclei per unit area, so the H~\textsc{i} and CO maps can be compared directly. The top panels show C1, the middle panels show C2, and the bottom panel shows C3; C3 was not detected in CO. Red circles in the bottom left-hand corner of each map show the beam size of the observations. The black horizontal line in the bottom right-hand corner of each map shows the physical scale at an assumed distance of 8.2 kpc, while arrows point toward the projected position of the Galactic Centre. The black dashed squares in the middle panels of C1 and C2 and in the map of C3 outline the field of view of the CO observations. The red dashed boxes in the middle panels of C1 and C2 outline the zoomed-in area of the right-hand panels. Contour levels for \hi{} are $ [0.5, 1.05, 1.6] \times 10^{20}$ H nuclei cm$^{-2}$ and contour levels for CO are $ [4.625, 7.75, 10.875, 14] \times 10^{20}$ H nuclei cm$^{-2}$. MeerKAT \hi{} data have a 4$\sigma_\mathrm{rms}$ mask, CO data have a 5$\sigma_\mathrm{rms,CO}$ mask.}
    \label{fig:cdmaps}
\end{figure*}

We derive neutral hydrogen column densities from the velocity-integrated brightness temperatures we measure with MeerKAT as:

\begin{equation}
\label{eqn:cd}
\nhi{} = 1.823 \times 10^{18} \int T_\mathrm{b} \: dv \: \: \: \mathrm{cm}^{-2}, 
\end{equation} 

\noindent where \nhi{} is the column density and $T_{\mathrm{b}}$ is the brightness temperature. The constant $1.823 \times 10^{18}$ assumes that the gas is optically thin, which is usually a good assumption for relatively low column density gas \citep{Dickey1990}. Using the Python package \emph{SpectralCube} \citep{Ginsburg2015}, we evaluate the velocity-integrated brightness temperature, $\int T_\mathrm{b} \: dv$. We show the resulting \hi\ column density maps within a given velocity range for each cloud in the left column of \autoref{fig:cdmaps}.

Given \nhi{}, the mass surface density of gas traced by \hi{} is 

\begin{equation}
\label{eqn:hisd}
\shi{} = \left( \frac{\nhi{}}{9.18 \times 10^{19} \: \mathrm{cm}^{-2}} \right) \mpc{}
\end{equation}

\noindent 
including helium; the numerical factor in the denominator assumes a hydrogen mass fraction $X=0.74$. The corresponding total mass for atomic gas is:
\begin{equation}
\label{eqn:masshi}
M_\mathrm{at} = 1.36 \: m_\mathrm{H} \: D^2 \int \nhi{}  \: d\Omega, 
\end{equation}
where the factor $1.36 = 1/X$ is to account for helium, $m_\mathrm{H}$ is the mass of the hydrogen atom, $D$ is the distance to the clouds, and the integral is over the solid angle, $\Omega$, subtended by the observations. We assume $D=8.2$ kpc (\citealt{Gravity2019}; see \aref{App:distance} for masses derived using kinematic distance estimates as described in \citealt{DiTeodoro2020}). We calculate the mass of C1 between velocity channels $148$ to $181$ \kms~within an area defined by $(l,b)=(-1.65^{\circ}\, \mathrm{to}\, -2.05^{\circ}, -3.71^{\circ} \,\mathrm{to}\, -4.10^{\circ})$. All detections $\geq \mathrm{4}\sigma_\mathrm{rms}$ are included, where $\sigma_\mathrm{rms}$ is the rms noise of the 0th-moment map. For C2, we consider all detections $\geq \mathrm{4}\sigma_\mathrm{rms}$ between velocity channels $236$ to $291$ \kms~within an area defined by $(l,b)=(-2.16^{\circ}\, \mathrm{to} \,-2.66^{\circ}, 5.37^{\circ} \,\mathrm{to}\, 5.87^{\circ})$. For C3, we consider detections $\geq \mathrm{4}\sigma_\mathrm{rms}$ between velocity channels $-121$ to $-77$ \kms{} within an area defined by $(l,b)=(12.87^{\circ}\, \mathrm{to}\, 13.78^{\circ}, 6.50^{\circ} \,\mathrm{to}\, 7.40^{\circ})$. To determine the velocity range of the clouds, we include all channels where there are either HI or CO detections. As no CO is detected in C3, the width was determined solely from the \hi{} emission.

We calculate $\sigma_{\mathrm{rms}}$ values individually for each cloud. To do so, we first calculate the channel standard deviation ($\sigma_{\mathrm{chn}}$) across a number of emission-free channels ($N_{\mathrm{chn}}$) and then calculate the $\sigma_{\mathrm{rms}}$ for the 0th-moment map as

\begin{equation}
 \label{eqn:std}
 \sigma_{\mathrm{rms}} = \sigma_{\mathrm{chn}} \Delta_{\mathrm{chw}} \sqrt{N_{\mathrm{chn}}/1.2}
 \end{equation}
 
\noindent where $\Delta_{\mathrm{chw}}$ is the channel width of the observations (5.5 \kms{}), and the factor 1.2 is the velocity spread approximation accounting for the fact that the channels are not independent. We then mask all pixels less than 4 $\sigma_{\mathrm{rms}}$ in the 0th-moment map.

\subsection{CO Observations}
\label{sec:coobs}
We compare the MeerKAT \hi\ data to observations of the $^{12}$CO(2$\rightarrow$1) emission line at 230.538 GHz carried out with the Atacama Pathfinder EXperiment (APEX). 
The CO data for C1 and C2 are taken from \citet{DiTeodoro2020}, and we refer to this work for data acquisition and reduction details.
C1 and C2 observations have an angular resolution of 28$^{\prime\prime}$ with a channel width of 0.25 \kms{}, covering a  $15^\prime \times 15^\prime$ field centred on each cloud. 
The rms noise ($\sigma_\mathrm{chn,CO}$) is 65 mK and 55 mK for C1 and C2, respectively, per 0.25 \kms{} channel. 
C3 was observed as a part of a new observational program with APEX (Di Teodoro et al., in prep.), using the new nFLASH heterodyne receiver and covering a bandwidth of 8 GHz with 61 kHz spectral channels (corresponding to 0.08 \kms). 
Two $15' \times 15'$ on-the-fly maps were observed for C3, centred on the densest regions of the \hi\ MeerKAT emission (see \autoref{fig:cdmaps} for an outline of the observed regions). 
Data reduction and imaging for C3 followed the standard procedure described in \citet{DiTeodoro2020} for C1 and C2.
The final rms noise for C3 is 82 mK in a 0.25 \kms\ channel with a spatial resolution of $28''$.
Unlike in C1 and C2, CO emission was not detected across any of the two fields of C3, not even after spectral smoothing to channel widths of 1 and 2 \kms.
Thus, C3 CO data just provide upper limits to the amount of molecular gas in the cloud. 

We derive the hydrogen column densities as traced by molecular hydrogen (\htwo) (see \autoref{fig:cdmaps}) from the CO observations as:
\begin{equation}
\label{eqn:Nh2}
N_\mathrm{H_2} = 2 \times 10^{20} \int T_\mathrm{b} \: dv \: \mathrm{cm}^{-2}, 
\end{equation} 
where all values are the same as in \autoref{eqn:cd} except for the $2 \times 10^{20}$ constant, which is the CO-to-\htwo{} conversion factor, $X_\mathrm{CO}$. This conversion factor is unknown within the Galaxy's nuclear wind, but \cite{DiTeodoro2020} use radiative-transfer and non-LTE excitation modelling with the \textsc{Despotic} code \citep{Krumholz2014} to conclude that $X_\mathrm{CO}=2 \times 10^{20}$ cm$^{-2}$ (K\ \kms{})$^{-1}$ is likely a lower limit for the nuclear wind. Given $N_{\mathrm{H_2}}$, we calculate the total mass surface density (including helium) as traced by \htwo{} as 
\begin{equation}
\label{eqn:h2sd}
\Sigma_{\mathrm{mol}} = \left( \frac{N_{\mathrm{H_2}}}{4.59 \times 10^{19} \mbox{cm}^{-2}} \right) \mpc{},
\end{equation}
where the numerical factor again assumes $X=0.74$, and the corresponding total mass as
\begin{equation}
\label{eqn:massh2}
M_\mathrm{mol} = 1.36 \: m_\mathrm{H_2} \: D^2 \int N_\mathrm{H_2}  \: d\Omega, 
\end{equation}
where $m_\mathrm{H_2}$ is the mass of molecular hydrogen. Note that, when comparing the \hi{} and \htwo{} data, to avoid confusion, we will always use either mass surface density or surface densities of H nuclei, $N_\mathrm{H} = 2 N_\mathrm{H_2}$.

All CO maps were observed by dividing the $15'\times15'$ fields into nine smaller $5'\times5'$ sub-maps. 
Each sub-map was observed under different weather conditions and with slightly different exposure times. 
As a result, the rms noise varies across each field. 

To account for this, we generate noise maps of each cube to produce an appropriate mask for the CO data. First, we calculated the rms noise of each voxel that does not contain cloud emission, producing a noise map, which is then multiplied by the channel width and the square root of the number of channels in the 0th-moment map. To create a signal-to-noise map (S/N), we divide the original moment map by the noise map. We then mask all pixels less than 5 in the S/N cube and use the masked S/N map to mask the original 0th-moment map. Note, the CO channels are independent of each other as they have been regridded from 0.08 \kms{} to 0.25 \kms{}.

\section{Analysis}
\label{sec:analysis}
We now analyse the \hi{} and CO data, starting with a discussion of their relative morphologies in \autoref{sec:morph} and then proceeding to a quantitative analysis of the molecular fraction of the clouds in \autoref{sec:hi2h2}.
When comparing the two datasets, we always bin the \hi{} data to the slightly coarser spatial resolution of the CO data, but we note that, even before binning, the CO and \hi{} data have very similar angular resolutions ($28^{\prime\prime}$ versus $24.9^{\prime\prime}\times 21.7^{\prime\prime}$), so binning effects are minimal.

\begin{figure}
	\includegraphics[width=8cm]{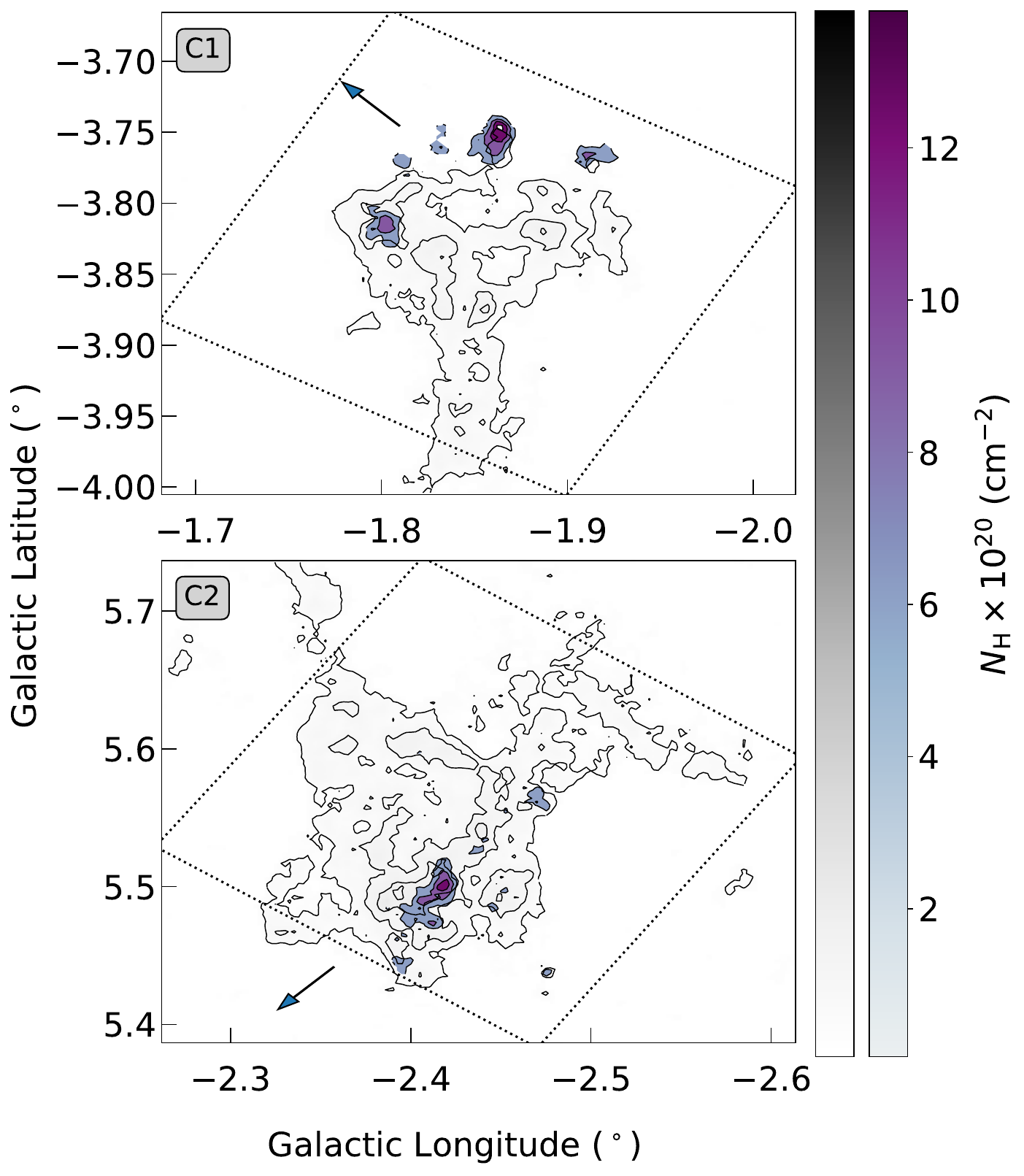}
    \caption{Same as \autoref{fig:cdmaps}, but now showing the CO data (blue-purple colours) superimposed on the \hi{} (greyscale). The top panel shows C1, whereas the bottom panel shows C2. The contour levels are the same as those in \autoref{fig:cdmaps}. Here $N_\mathrm{H} = 2 N_\mathrm{H_2}$.}
    \label{fig:overlays}
\end{figure}
\subsection{Morphological comparison}
\label{sec:morph}
 
\autoref{fig:cdmaps} shows the column density maps of the clouds in \hi{} and CO separately, and \autoref{fig:overlays} shows the CO superimposed on the \hi{} for C1 and C2; in the latter figure, we have rebinned the \hi{} data to the same spatial grid as the CO. For C1, the CO emission consists of five fragmented clumps aligned almost perpendicular to the direction of the GC. The \hi{} trails behind the compact CO clumps as the distance from the GC increases. The \hi{} produces a tail-like morphology, while the CO appears to be concentrated near the head of the cloud. For C2, the CO is central to the \hi{}, aligned parallel to the direction of the GC. The CO and \hi{} peaks nearly coincide spatially, and both species trail off in column density as the distance from the GC increases. The CO appears to be at the core of the cloud, surrounded by the \hi{}. 

\begin{figure}
	\includegraphics[width=8.6cm]{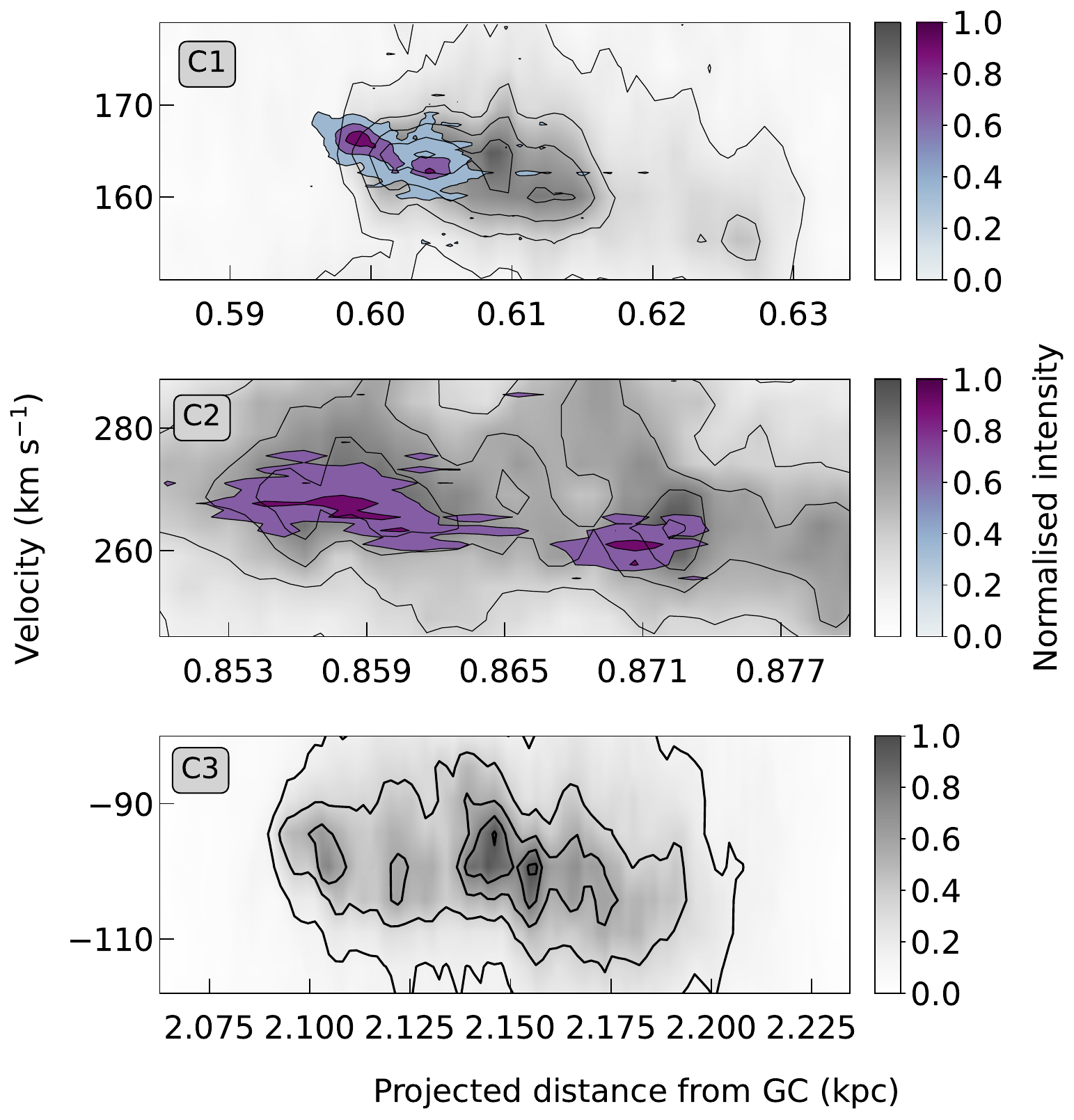}
    \caption{Position-velocity maps of the CO (blue-purple colours) and \hi{} (greyscale) observations where velocity is the V$_{LSR}$. The top panel shows C1, the middle panel shows C2, and the bottom panel shows C3. The position is measured by projected distance from the GC, computed as discussed in \autoref{sec:morph}, and we compute the intensity at a given position-velocity point by integrating the observed PPV cube along the direction transverse to the GC. Both \hi{} and CO data are normalised to have a maximum of unity in the PV diagram. \hi{} contour levels are $[0.15, 0.35, 0.55, 0.75, 0.95]$, CO contour levels are $[0.2, 0.5, 0.8, 1.]$.}
    \label{fig:pvmaps}
\end{figure}

We show all three clouds' position-velocity (PV) maps in \autoref{fig:pvmaps} where velocity is the V$_{LSR}$. This figure measures positions in terms of projected distances from the GC, which we calculate assuming that both the cloud and the GC are at the same distance $D=8.2$ kpc from the Sun. To construct the PV map, we integrate all emissions in the direction transverse to the GC and normalise all intensities to a common scale so that we can easily compare the H\textsc{i} and CO data.

For C1, the CO lies closer to the GC than the \hi{}, consistent with what we see in the column density maps, whereas for C2, the CO coincides spatially and spectrally with the \hi{}. Further, \autoref{fig:pvmaps} reveals that the velocity structures between the \hi{} and CO are similar for C1 and C2. In both, there is a clear gradient whereby as the projected distance from the GC increases, there is a $\sim$ 5-10 \kms{} decrease in velocity with a $\sim -1$ \kms{} / pc gradient. For C3, whilst there is no CO to compare with the \hi{}, the \hi{} shows a velocity gradient similar to that seen in C1 and C2. 

\begin{figure}
	\includegraphics[width=8.6cm]{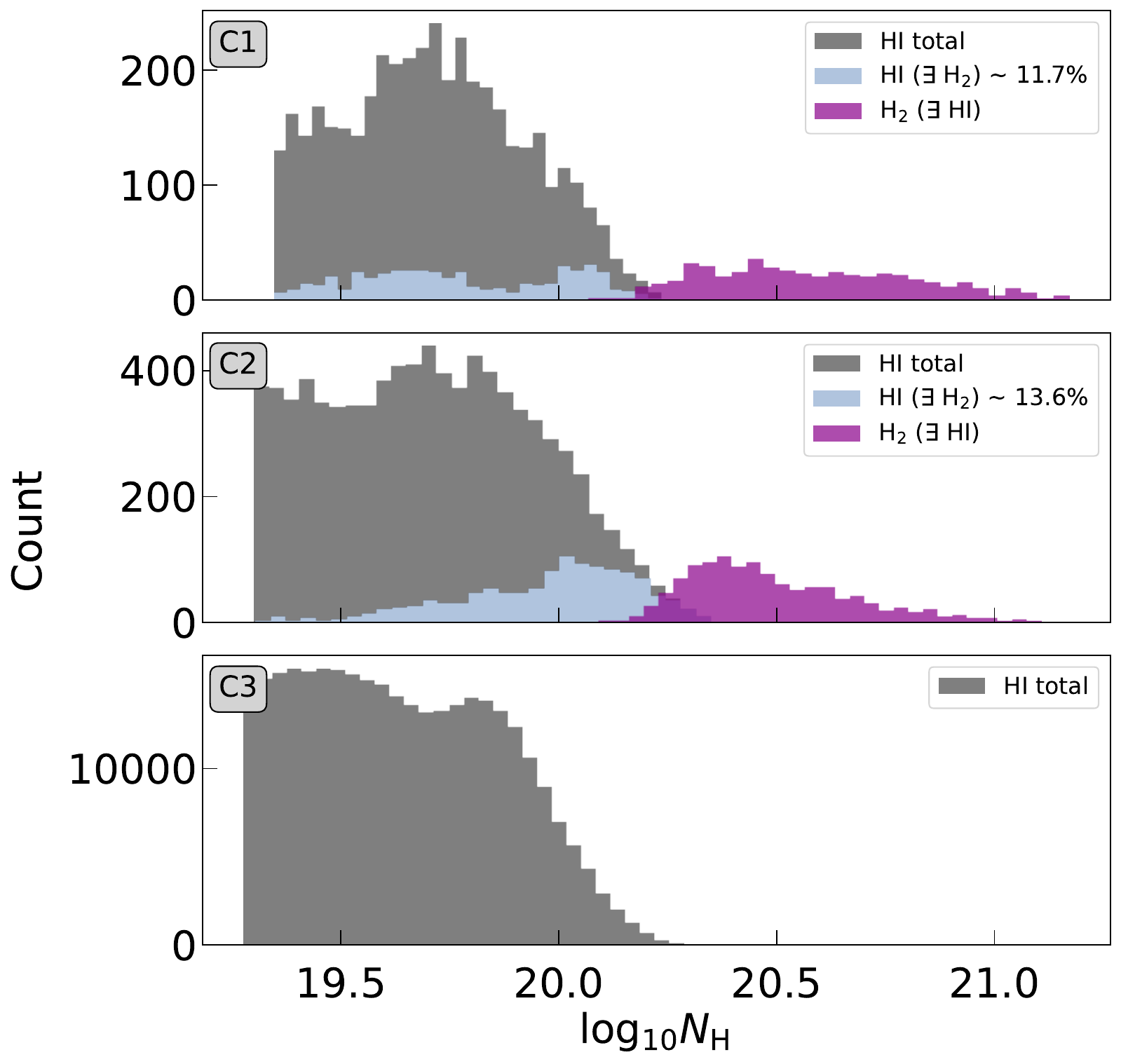}
    \caption{Histograms of pixel counts for total \hi{} column densities in grey, \hi{} columns limited to pixels where there is also a detection in CO in light blue, and \htwo{} columns where there is also a detection in \hi{} in purple. The fraction of \hi{}-detected pixels, where there is also a CO detection, is indicated in the legend for the C1 and C2 panels. Here, \htwo{} is in terms of H nuclei, $N_\mathrm{H} = 2 N_\mathrm{H_2}$.}
    \label{fig:hist}
\end{figure}

We further investigate the cloud morphologies by comparing the probability distribution functions (PDFs) of \hi{} and \htwo{} in \autoref{fig:hist}, distinguishing between the total \hi{} distribution and the distribution considering only those pixels for which we also detect CO. Here we notice a significant difference between the clouds: while C1 and C2 have similar column density PDFs overall, for C1 sight lines where there are detections in both \hi{} and CO cover most of the \hi{} column density range; except at the highest \hi{} columns, there is little correlation between the \hi{} column density in a pixel and the probability that it will harbour detectable CO emission. For C2, by contrast, \hi{} pixels where we detect CO are much more heavily biased toward the highest \hi{} columns, consistent with the visual impression in \autoref{fig:overlays}. For C3, there are no CO detections, and the \hi{} column density PDF is shifted to lower values on average than those found in C1 and C2. Nonetheless, C3 contains significant numbers of pixels at \hi{} column densities similar to those at which C1 and C2 show significant CO detection fractions. Thus the overall conclusion to draw from \autoref{fig:hist} is that the \hi{}-\htwo{} relationship is substantially different in all three clouds.

\subsection{\hi{}-to-\htwo{} ratio and molecular fraction}
\label{sec:hi2h2}

\begin{figure*}
	\includegraphics[width=16cm]{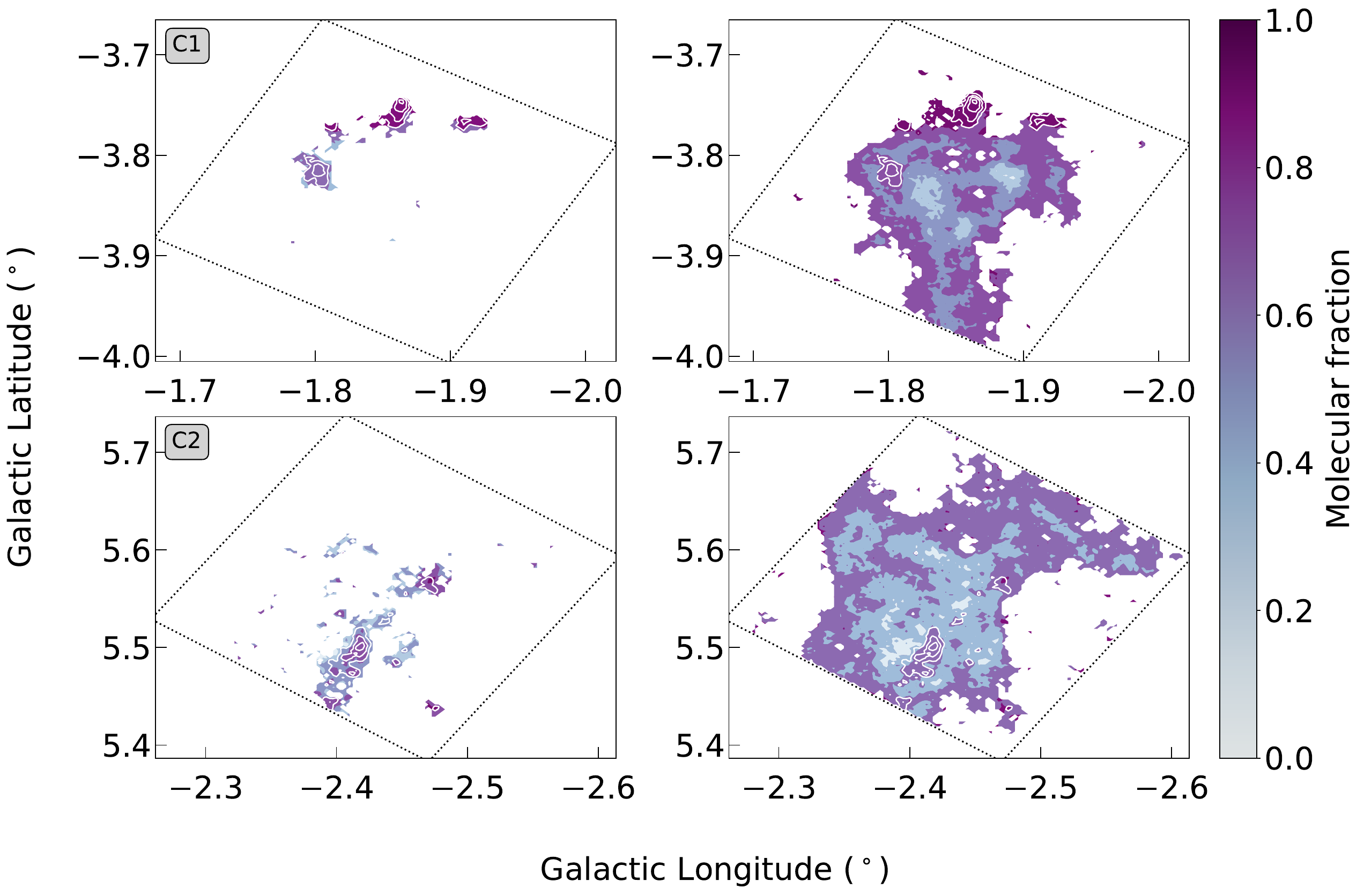}
    \caption{Molecular fraction maps with C1 in the upper panels and C2 in the lower panels. The left panels show the molecular fraction only when there is a detection in both \hi{} and CO, while the right panels also include upper limits on the molecular fraction derived from upper limits on the CO in those pixels where CO is not detected but \hi{} is; the CO upper limits correspond to our masking level of 5 times the (position-dependent) rms noise level in each pixel. White contours show the CO data at the same contour levels as those used in \autoref{fig:cdmaps}. The black dotted box is the CO observation field of view. }
    \label{fig:molfrac}
\end{figure*}

\begin{figure}
	\includegraphics[width=8.6cm]{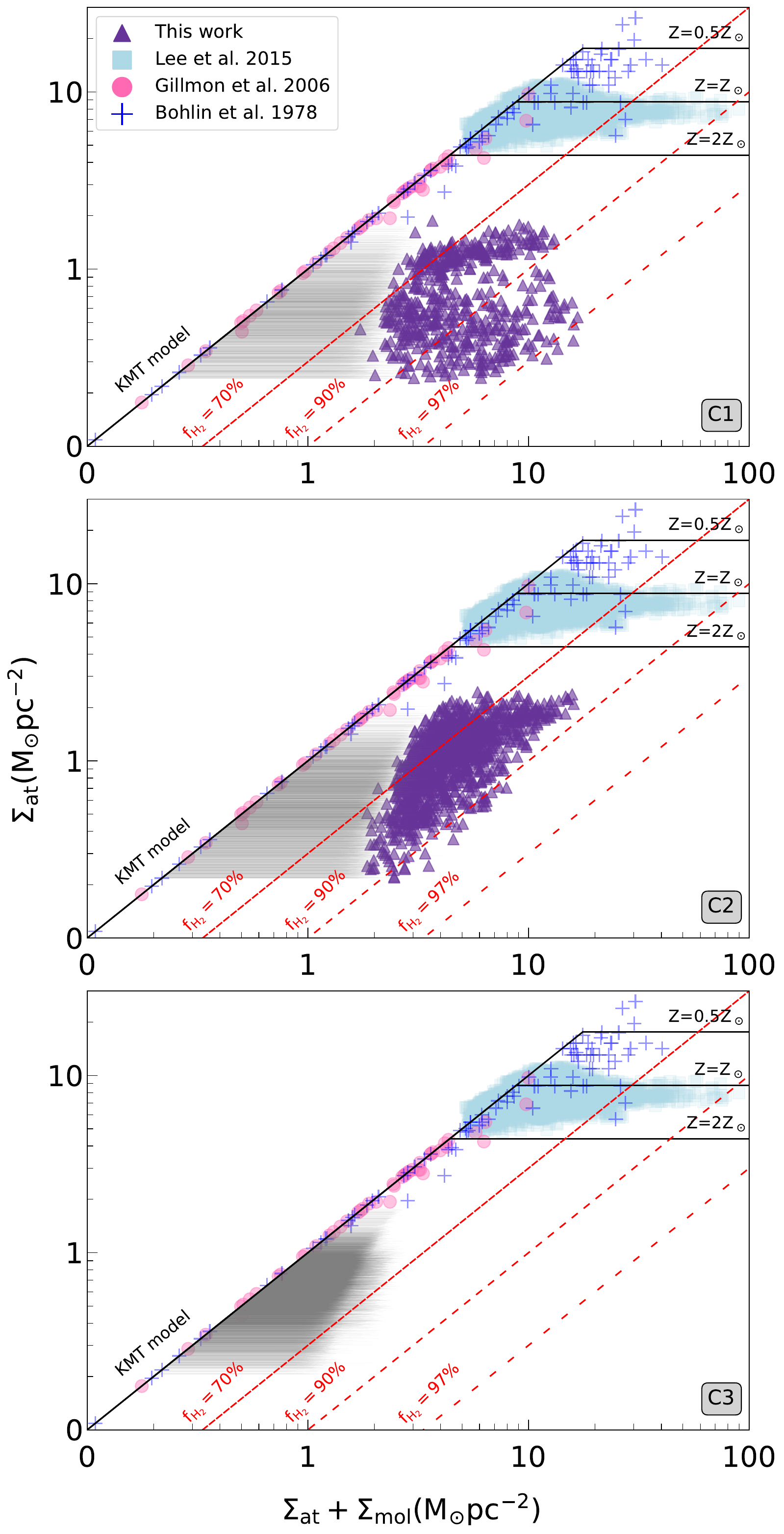}
    \caption{\shi{} vs. \shi{} + \shii{} for C1 (top), C2 (middle) and C3 (bottom). Purple triangles are detections in both H~\textsc{i} and CO, while the horizontal grey lines are the upper and lower limits of \shi{} + \shii{} derived by taking the minimum H$_2$ column to be 0 and the maximum to be the H$_2$ column corresponding to the CO detection threshold; this is 5$\sigma_\mathrm{rms, CO}$. Black solid lines show the \hi-\htwo~relation predicted by the KMT model \citep{Krumholz2008, Krumholz2009, McKee2010} for half-solar, solar and double solar metallicities; the diagonal portion of this line has slope unity and corresponds to a purely \hi~composition, with no \htwo. Red, dashed lines show different molecular fractions ranging from f$_{\mathrm{H}_2} = 70\% - 97\%$. Light blue squares show measurements from \citet{Lee2015} for five dark and star-forming regions in the Perseus molecular cloud in the Milky Way plane, pink circles show results from the FUSE survey of high-latitude \htwo{} regions from \citet{Gillmon2006} and dark blue crosses show results from Copernicus observations \citep{Bohlin1978}.}
    \label{fig:surfdens}
\end{figure}

We next investigate the molecular fractions (M$_\mathrm{mol}$/(M$_\mathrm{at}$+M$_\mathrm{mol})$) of the clouds by showing this fraction as a function of position in \autoref{fig:molfrac}, and by plotting \shi{} vs. \shi{} + \shii{} pixel-by-pixel in \autoref{fig:surfdens}. 

Surface densities are derived from the masked maps as described in \autoref{sec:hiobs} and \autoref{sec:coobs}. The left panels of \autoref{fig:molfrac} show the molecular fraction only when there is a detection in both \hi{} and CO, while the right panels are the same map but present upper limits for pixels where there is a non-detection in CO but a detection in \hi{}. We calculate the upper limit of CO using the position-dependent masking level computed as described in \autoref{sec:coobs}. We see that for the two clouds containing a CO component, the molecular fraction is high ($\sim 1$) over much of the area where there is a detection in both \hi{} and CO. When considering the upper limits for CO, $99\%$ of C1 and C2 may have a molecular fraction $\geq 50\%$. 

Similarly, in \autoref{fig:surfdens} the purple triangles show only pixels when there is a detection in both CO and \hi{}, while horizontal grey lines show pixels where we have an \hi~detection but only an upper limit for CO, with the limits for H$_2$ derived from the CO noise maps; the grey lines show a range from an H$_2$ column of 0 (i.e., $\shi{} + \shii{} = \shi{}$) to a total column density derived by adding $\shi{}$ the H$_2$ column corresponding to the level below which CO is masked; this is 5$\sigma_{\mathrm{rms,CO}}$. For C3, since there is no observed CO emission, we only show upper limits in the bottom panel of \autoref{fig:surfdens}.

For comparison, we also show \shi{} vs. \shi{} + \shii{} for five dark and star-forming regions in the Perseus molecular cloud in the Solar neighbourhood observed by \citet{Lee2015} (light blue squares), results from the Far Ultraviolet Spectroscopic Explorer (FUSE) survey of high-latitude, interstellar \htwo{} as presented in \cite{Gillmon2006} (pink circles), results from the Copernicus satellite interstellar Lyman $\alpha$ lines as presented in \cite{Bohlin1978} (dark blue crosses), along with the theoretical relation describing the equilibrium \hi-H$_2$ phase balance \citep{Krumholz2008, Krumholz2009, McKee2010} (black line) for varying metallicities (Z=0.5Z$_\odot$, Z$_\odot$ and 2Z$_\odot$). We refer to this equilibrium line as the KMT (Krumholz-McKee-Tumlinson) model. We also show orange lines indicating where the sight lines have molecular fractions (f$_\mathrm{H_2}$) of $70\%$, $90\%$ and $97\%$.

For Solar metallicity (Z=Z$_\odot$), the KMT model predicts that \hi~converts to H$_2$, thereby setting a maximum \hi~column, at $\Sigma_\mathrm{at} = 8.8$ M$_\odot$ pc$^{-2}$. The most important thing to note from this figure is that, for C1 and C2, all \shi{} values are in the range $\sim$ 0.3-3 \mpc{}, well below the $\sim$ 10 \mpc{} level that the KMT model predicts is required to shield molecular gas against photodissociation for Solar metallicity, and which is observed in the Perseus cloud, the high-latitude sight lines, and the interstellar Lyman $\alpha$ lines. Despite this, we detect appreciable columns of H$_2$, with H$_2$ dominating the total column on at least some lines of sight.

Thus the fraction of \htwo~relative to \hi~in these clouds is very different from the atomic to molecular fraction observed in the Galactic plane, where $\sim 1$ \mpc{} columns of \hi~are essentially never observed to host significant quantities of H$_2$, as shown in \autoref{fig:surfdens}. The presence of molecular material with low columns of \hi{} implies that $N_\mathrm{HI} \approx 20^{21}$ cm$^{-2}$ is not the sole determinant of the presence of \htwo{} as expected for gas in chemical equilibrium \citep{Krumholz2009} and as observed \citep{Liszt2023} in the Galaxy.

We emphasise that this result is \textit{not} dependent on our (highly uncertain) choice of $X_\mathrm{CO}$. Using a different value of $X_\mathrm{CO}$ would shift the cloud of observed points left or right in \autoref{fig:surfdens}, and thus could not bring the data close to the locus occupied by the Perseus, \cite{Gillmon2006}, or \cite{Bohlin1978} points. Moreover, given that many of the points have $f_\mathrm{H_2} > 90\%$, shifting them far enough to the left so that they lie close to the KMT model line (indicating a purely atomic composition) and the \hi{}-dominated FUSE and Copernicus points would require a value of $X_\mathrm{CO}$ more than an order of magnitude smaller than our fiducial estimate, and close to the absolute limit for optically thin CO emission $X_\mathrm{CO} \approx 10^{19}$ cm$^{-2}$ / ($\mbox{K km s}^{-1}$) \citep{Bolatto2013}.

Nor could the result be explained by GC gas having super-Solar metallicity. Metallicity clearly does affect the \hi{} column at which the transition to \htwo{} occurs as demonstrated by the \cite{Bohlin1978} points which have slightly sub-solar metallicities, accounting for the offset from the Z=Z$_\odot$ line. However, the KMT model predicts that, in order to form \htwo{} at \hi{} column densities of $\sim 1$ \mpc{} as we observe, the dust abundance would have to be $\approx 10\times$ the Solar neighbourhood value, implausibly high, and even in the most metal-rich external galaxies the \hi{} to \htwo{} transition is observed to begin only at \hi{} surface densities $\sim 5$ \mpc{} \citep{Wong2013, Schruba2018}. For a realistic GC metallicity of $Z \lesssim 2Z_\odot$, the observed points still lie far from the equilibrium KMT model prediction.

\section{Discussion}
\label{sec:disc}
In this discussion, we investigate the likely origins of both atomic and molecular gas entrained in the Milky Way's nuclear wind by considering the morphologies of the two observed species in \autoref{sec:discmorph} and the chemical states of each cloud in \autoref{sec:discchem}. We finish this section by considering the lifetime of a cloud in the GWs in \autoref{sec:disctime}.

\subsection{Morphologies}
\label{sec:discmorph}
The two clouds where we detect both \hi\ and CO show contrasting morphologies, potentially suggesting we observe clouds of different origins or, as suggested in \cite{DiTeodoro2020}, different evolutionary stages of the same process. If the molecular gas is disassociating into atomic hydrogen, we could expect to see the \hi{} surround the molecular gas as we see in molecular clouds in the disc \citep[e.g.][]{Stanimirovic2014}. However, it is also conceivable that in the presence of a strong wind, the \hi{} could trail behind the \htwo{} due to stripping effects -- that is, \hi\ may ``boil'' off the surface of the molecular cloud but then be swept back by the ram pressure of the unseen hot gas that drives the wind. We see both of these morphologies across C1 and C2 (see \autoref{fig:overlays}, \autoref{fig:hist}).\footnote{The difference between the two clouds may also be a result of projection effects: \citet{DiTeodoro2018} present a kinematic model of the wind that predicts that C1 lies almost directly transverse to the GC from our perspective, while C2 is a similar projected distance from the GC but is several kpc closer to us in the radial direction. Thus the radial vector from each cloud to the GC lies nearly in the plane of the sky for C1, but is significantly tilted relative to the plane of the sky for C2.} In the early stages of cloud evaporation, \hi{} would be a surface phenomenon unrelated to the depth of the cloud, as appears to be the case for C1 in \autoref{fig:hist}. Later on, when the disassociation front has broken up much of the \htwo{}, fresh \hi{} would continue to be produced from the remaining dense \htwo{} regions, as appears to be the case for C2 in \autoref{fig:hist}. Finally, in the late stages of cloud evaporation, all \htwo{} would have dissociated into \hi{}, as appears to be the case in C3. 

Alternatively, if the \hi\ is compressing into molecular gas, we expect this to occur at the central, densest regions of the cloud. As the molecular gas is positioned along the inner regions of C2 and towards the GC with \hi{} trailing behind in C1, either scenario (or a combination of the two) could be taking place. Given the lack of molecular matter in C3 and its greater distance from the GC compared to C1 and C2 (\autoref{fig:pvmaps}), it is possible it once had a molecular component, but it has since all dissociated into \hi{}. Lastly, when considering the third scenario presented in \autoref{sec:intro}, it is difficult to predict the expected geometry for a cool cloud that has condensed out of the rapidly cooling hot phase of GWs. Thus a cloud's morphology alone cannot disentangle these scenarios. 
\subsection{Chemical state}
\label{sec:discchem}
The chemical state of the clouds is more informative. As discussed in \autoref{sec:hi2h2} and shown in \autoref{fig:molfrac}, there are significant portions of C1 and C2 by both mass and area where the molecular fraction is close to 1, and considering our sensitivity limits, there may be additional undetected regions of moderately molecular ($f_{\mathrm{H}_2} \sim 0.3-0.5$) material. More importantly, as shown in \autoref{fig:surfdens}, this high H$_2$ fraction is greater than we would expect given the low \hi{} column densities, based on either theoretical models or on the measured \hi{}-\htwo{} relationship in the Galactic plane. While high molecular fractions in CO-detected pixels are not surprising since low levels of CO column densities are either undetectable or have been masked, it is nonetheless significant that we detect CO at all, indicating the presence of considerable columns of \htwo{} at levels of \hi{} where normally there would not be any \citep[e.g.][]{Krumholz2009, Liszt2023}.

This over-abundance in \htwo{} makes a compression-driven H$_2$ formation scenario unlikely. While it is easy to understand how the \htwo{} chemistry could be out of equilibrium, given the relatively long equilibration timescales (a point to which we return momentarily), if the system were transitioning from \hi{}-dominated to beginning to form \htwo{}, we would expect the system to be out of equilibrium in the direction of too little \htwo{}, not too much as we observe. Quantitatively, for evolutionary times $t$ smaller than the equilibrium time scale $t_\mathrm{eq}$, we expect the molecular fraction $\Sigma_\mathrm{mol} / (\Sigma_\mathrm{at} + \Sigma_\mathrm{mol})$ to be intermediate between the initial state which exists at $t \ll t_\mathrm{eq}$ and the equilibrium state achieved at $t \gg t_\mathrm{eq}$. Thus if the initial state contains no significant amount of \htwo{}, $\Sigma_\mathrm{mol} / (\Sigma_\mathrm{at} + \Sigma_\mathrm{mol}) \approx 0$, and the final equilibrium has a non-zero \htwo{} fraction, as posited in the compression scenario, then we would expect that at all times the value of $\Sigma_\mathrm{mol} / (\Sigma_\mathrm{at} + \Sigma_\mathrm{mol})$ would be intermediate between zero and the equilibrium value; at no point would we expect to find the system in a state where it \textit{exceeds} the equilibrium value, which is the state that we actually observe. The same argument implies that it is also unlikely that the clouds are a result of condensation of cool material out of a hot phase.

By contrast, the observed \hi{} surface densities and over-abundance of \htwo{} are consistent with our first scenario, where a stripping of \hi{} from the envelope of a pre-existing molecular cloud leaves behind too little \hi{} to shield the \htwo{} from the Galaxy's dissociating radiation field, and the \htwo{} is in the process of converting to \hi{}. This scenario also explains the non-detection of CO emission from C3: this cloud is significantly further from the GC in projection than C1 or C2, and thus has likely had more time to undergo photodissociation than the other two clouds. The photodissociation process has been completed for this cloud, leaving behind only \hi{}. Consistent with this picture, in \autoref{fig:surfdens} we also observe that the cloud closest to the disc (C1) has a wider spread of points in the direction below and to the right of the equilibrium line, implying a higher \htwo{} abundance, whilst C2 (which is further from the disc) has points that, while still indicating significant amounts of \htwo{}, lie closer to the line. In other words, as clouds move further from the disc, the points on the $\shi{}$ vs.~$\shi{} + \shii{}$ plot migrate closer towards the theoretical equilibrium line as the clouds' molecular components dissociate.

\subsection{The lifetime of molecular clouds in the Milky Way's nuclear wind}
\label{sec:disctime}
In the dissociation scenario, we can estimate the lifetimes of molecular clouds in the wind from theory and check that the resulting predictions are consistent with the timescales inferred from cloud kinematics. For the theoretical estimate, we follow the general approach of the KMT model \citep{Krumholz2008, Krumholz2009, McKee2010}:\footnote{We note that in addition to the photodissociation process we consider here, in principle, H$_2$ could be collisionally-dissociated as well; this could occur if either turbulent mixing or conduction raise the temperature in the H$_2$ to several thousand K, hot enough for collisional dissociation to be significant. However, this seems unlikely to contribute to the \hi{}-\htwo{} balance directly because the similarity between the \htwo{} binding energy (4.5 eV) and the \hi{} ionisation energy (13.6 eV) means that there is only a very narrow range of temperature ($\approx 0.5 - 1.5\times 10^4$ K) where collisions will dissociate \htwo{} but not also ionize \hi{}. Below this range, collisional dissociation is unimportant, and above it, collisions will rapidly convert both \htwo{} and \hi{} to H$^+$. Thus for collisions to be an effective means of converting \htwo{} to \hi{} requires fine-tuning of the temperature.} we first note that the flux of dissociating Lyman-Werner band photons on the surface of a cloud exposed to the full, unshielded interstellar radiation field (ISRF) is $F^* = 2.1\times 10^7 \chi$ photons cm$^2$ s$^{-1}$, where $\chi$ is the ISRF strength normalised to the Solar neighbourhood value. If this flux falls on an optically thick cloud of fully molecular gas, the very large ratio of the H$_2$ opacity to the dust opacity ensures that almost all the photons will be absorbed by H$_2$ molecules, and a fraction $f_\mathrm{diss}\approx 0.1$ of these absorptions will result in the dissociation of the molecule. This will convert \htwo{}-dominated material to \hi{}-dominated at a rate (expressed as mass converted per unit area per unit time)
\begin{equation}
    \dot{\Sigma}_\mathrm{mol\to at} = 2 f_\mathrm{diss} F^* \frac{m_\mathrm{H}}{X} = 1.4\chi\;\mathrm{M}_\odot\mbox{ pc}^{-2}\mbox{ Myr}^{-1}
\end{equation}
where the factor of two accounts for there being two H nuclei per H$_2$ molecule. Thus the characteristic time required to dissociate a molecular cloud of initial surface density $\Sigma_\mathrm{H_2}$ is
\begin{equation}
    t_\mathrm{diss} = \frac{\Sigma_\mathrm{mol}}{\dot{\Sigma}_\mathrm{mol\to at}} = 7.0 \left(\frac{\Sigma_\mathrm{mol}/\chi}{10\;\mathrm{M}_\odot\;\mathrm{pc}^{-2}}\right)\;\mathrm{Myr}.
\end{equation}

This estimate carries significant uncertainties -- most obviously the strength of the ISRF $\chi$, but also that, as dissociation proceeds and a shielding layer of H~\textsc{i}-dominated material builds up, the conversion rate will fall because dissociations will be offset by H$_2$ formation in the shielding layer and because potentially dissociating photons will be absorbed by dust grains while they transit the shielding zone. The actual dissociation time will depend on the rate at which hydrodynamic processes strip away this shielding layer, exposing more unshielded H$_2$; in the absence of such stripping, a H$_2$ cloud with a surface density $\Sigma_\mathrm{mol}\gtrsim 10$ M$_\odot$ pc$^{-2}$ will eventually produce a shielding layer that will fully balance dissociation, and the remaining H$_2$ will survive indefinitely. Nonetheless, this rough estimate suggests that the characteristic survival time of H$_2$-dominated clouds whose shielding layers are being removed is a few to $\sim 10$ Myr.

This timescale lines up well with the kinematically-estimated ages of our observed clouds. Applying the model of \citet{Lockman2020} to our three clouds yields kinematic age estimates of $\approx 3.5$ Myr, $\approx 7$ Myr, and $\approx 8.3$ Myr with corresponding distances from the GC of 0.8 kpc, 1.8 kpc and 2.1 kpc for C1, C2, and C3, respectively. These ages are entirely consistent with our theoretical expectation that molecular clouds injected into the wind without adequate shielding should retain significant H$_2$ fractions for times of a few to 10 Myr, and favour a picture where C1 and C2 are out of equilibrium and are partway through the process of converting to \hi, while C3, which is somewhat older, is further along the path towards an equilibrium chemical configuration with negligible \htwo{}. Thus observations as well a theory support a $\sim \mbox{few}-10$ Myr lifetime for molecular clouds in the wind.

\section{Conclusions}
\label{sec:conc}
We present new high-resolution MeerKAT observations of three atomic clouds entrained in the Milky Way galactic wind. We combine these with APEX CO observations at closely matched resolutions to characterise the clouds' molecular content. We use these data to compare the atomic and molecular morphologies, and to determine spatially-resolved molecular fractions. Of the three clouds, the one closest to the Galactic Centre in projection (C1) has its molecular component positioned towards the Galactic Centre and the atomic component trailing behind, the intermediate-distance cloud (C2) has its molecular component positioned within the body of the \hi{} near the \hi{} column density maximum, and the cloud furthest from the Galactic Centre (C3) contains no detected molecular component at all. All three clouds show a velocity gradient of $\sim -1$ \kms{} / pc with distance from the Galactic Centre. 
We find molecular fractions by mass in C1 and C2 that are quite high, $\gtrsim 50\%$ and $\gtrsim 40\%$, respectively.
In many pixels where we detect H$_2$,  molecular material forms a significant fraction of the total. 

Most importantly, we find that the \hi{}-\htwo{} chemistry is out of equilibrium, and that, as a result, the GC clouds lie in a portion of \hi{}-\htwo{} parameter space that is unpopulated in the Galactic disc. The \hi{} surface density \shi{} is $0.3-3$ \mpc{} despite the presence of substantial quantities of H$_2$, much lower than $\sim 10$ \mpc{} \hi{} columns observed around molecular clouds in the disc \citep[e.g.][]{Lee2015, Liszt2023}, and suggested to be necessary to shield H$_2$ by the KMT model \citep{Krumholz2008, Krumholz2009, McKee2010}. This implies that there is insufficient \hi{} column density to shield the molecular gas against dissociation. As a result of this disequilibrium, we conclude that the three clouds likely originate from molecular clouds in the Galactic disc that were entrained in the Galactic wind and are photodissociated into atomic hydrogen by the radiation field in the winds. This picture is also consistent with the non-detection of C3 in CO, since this is the dynamically oldest of the three clouds and has had the longest to photodissociate. Comparing the kinematic ages with theoretical predictions for the \htwo{} dissociation rate, we conclude that the data are consistent with a picture where the wind consists primarily of molecular gas entrained into the outflow, which has a characteristic survival time of  $\sim 10$ Myr after it is entrained before it dissociates into \hi.

\section*{Acknowledgements}
The MeerKAT telescope is operated by the South African Radio Astronomy Observatory, which is a facility of the National Research Foundation, an agency of the Department of Science and Innovation. The authors acknowledge and thank Min-Young Lee and Sne\v{z}ana Stanimirovi\'{c} for providing the data from \cite{Lee2015} in electronic form. MRK acknowledges support from the Australian Research Council through Laureate Fellowship FL220100020. EDT was supported by the European Research Council (ERC) under grant agreement no. 10104075. NM-G is the recipient of an Australian Research Council Australian Laureate Fellowship (project number FL210100039) funded by the Australian Government.
The MeerKAT telescope is operated by the South African Radio Astronomy Observatory, which is a facility of the National Research Foundation, an agency of the Department of Science and Innovation.
The Green Bank Observatory and the Green Bank Telescope are facilities of the U.S. National Science Foundation, operated by Associated Universities, Inc.

\section*{Data Availability}
The data underlying this article along with a general implementation of the code used to process the data cubes and produce all of the figures in this paper is available via Zenodo at https://doi.org/10.5281/zenodo.8060960.



\bibliographystyle{mnras}
\bibliography{mnras_template} 

\begin{thebibliography}{}
\makeatletter
\relax
\def\mn@urlcharsother{\let\do\@makeother \do\$\do\&\do\#\do\^\do\_\do\%\do\~}
\def\mn@doi{\begingroup\mn@urlcharsother \@ifnextchar [ {\mn@doi@}
  {\mn@doi@[]}}
\def\mn@doi@[#1]#2{\def\@tempa{#1}\ifx\@tempa\@empty \href
  {http://dx.doi.org/#2} {doi:#2}\else \href {http://dx.doi.org/#2} {#1}\fi
  \endgroup}
\def\mn@eprint#1#2{\mn@eprint@#1:#2::\@nil}
\def\mn@eprint@arXiv#1{\href {http://arxiv.org/abs/#1} {{\tt arXiv:#1}}}
\def\mn@eprint@dblp#1{\href {http://dblp.uni-trier.de/rec/bibtex/#1.xml}
  {dblp:#1}}
\def\mn@eprint@#1:#2:#3:#4\@nil{\def\@tempa {#1}\def\@tempb {#2}\def\@tempc
  {#3}\ifx \@tempc \@empty \let \@tempc \@tempb \let \@tempb \@tempa \fi \ifx
  \@tempb \@empty \def\@tempb {arXiv}\fi \@ifundefined
  {mn@eprint@\@tempb}{\@tempb:\@tempc}{\expandafter \expandafter \csname
  mn@eprint@\@tempb\endcsname \expandafter{\@tempc}}}

\bibitem[\protect\citeauthoryear{{Andersson}, {Roger}  \&
  {Wannier}}{{Andersson} et~al.}{1992}]{Andersson1992}
{Andersson} B.~G.,  {Roger} R.~S.,   {Wannier} P.~G.,  1992, \aap, \href
  {https://ui.adsabs.harvard.edu/abs/1992A&A...260..355A} {260, 355}

\bibitem[\protect\citeauthoryear{{Bland-Hawthorn} \& {Cohen}}{{Bland-Hawthorn}
  \& {Cohen}}{2003}]{BlandHawthorn2003}
{Bland-Hawthorn} J.,  {Cohen} M.,  2003, \mn@doi [\apj] {10.1086/344573}, \href
  {https://ui.adsabs.harvard.edu/abs/2003ApJ...582..246B} {582, 246}

\bibitem[\protect\citeauthoryear{{Bohlin}, {Savage}  \& {Drake}}{{Bohlin}
  et~al.}{1978}]{Bohlin1978}
{Bohlin} R.~C.,  {Savage} B.~D.,   {Drake} J.~F.,  1978, \mn@doi [\apj]
  {10.1086/156357}, \href
  {https://ui.adsabs.harvard.edu/abs/1978ApJ...224..132B} {224, 132}

\bibitem[\protect\citeauthoryear{{Bolatto}, {Wolfire}  \& {Leroy}}{{Bolatto}
  et~al.}{2013}]{Bolatto2013}
{Bolatto} A.~D.,  {Wolfire} M.,   {Leroy} A.~K.,  2013, \mn@doi [\araa]
  {10.1146/annurev-astro-082812-140944}, \href
  {https://ui.adsabs.harvard.edu/abs/2013ARA&A..51..207B} {51, 207}

\bibitem[\protect\citeauthoryear{{Bregman}}{{Bregman}}{1980}]{Bregman1980}
{Bregman} J.~N.,  1980, \mn@doi [\apj] {10.1086/157867}, \href
  {https://ui.adsabs.harvard.edu/abs/1980ApJ...237..280B} {237, 280}

\bibitem[\protect\citeauthoryear{{Carretti} et~al.,}{{Carretti}
  et~al.}{2013}]{Carretti2013}
{Carretti} E.,  et~al., 2013, \mn@doi [\nat] {10.1038/nature11734}, \href
  {https://ui.adsabs.harvard.edu/abs/2013Natur.493...66C} {493, 66}

\bibitem[\protect\citeauthoryear{{Cashman} et~al.,}{{Cashman}
  et~al.}{2021}]{Cashman2021}
{Cashman} F.~H.,  et~al., 2021, \mn@doi [\apjl] {10.3847/2041-8213/ac3cbc},
  \href {https://ui.adsabs.harvard.edu/abs/2021ApJ...923L..11C} {923, L11}

\bibitem[\protect\citeauthoryear{{Crocker}, {Bicknell}, {Taylor}  \&
  {Carretti}}{{Crocker} et~al.}{2015}]{Crocker2015}
{Crocker} R.~M.,  {Bicknell} G.~V.,  {Taylor} A.~M.,   {Carretti} E.,  2015,
  \mn@doi [\apj] {10.1088/0004-637X/808/2/107}, \href
  {https://ui.adsabs.harvard.edu/abs/2015ApJ...808..107C} {808, 107}

\bibitem[\protect\citeauthoryear{{Di Teodoro}, {McClure-Griffiths}, {Lockman},
  {Denbo}, {Endsley}, {Ford}  \& {Harrington}}{{Di Teodoro}
  et~al.}{2018}]{DiTeodoro2018}
{Di Teodoro} E.~M.,  {McClure-Griffiths} N.~M.,  {Lockman} F.~J.,  {Denbo}
  S.~R.,  {Endsley} R.,  {Ford} H.~A.,   {Harrington} K.,  2018, \mn@doi [\apj]
  {10.3847/1538-4357/aaad6a}, \href
  {https://ui.adsabs.harvard.edu/abs/2018ApJ...855...33D} {855, 33}

\bibitem[\protect\citeauthoryear{{Di Teodoro}, {McClure-Griffiths}, {Lockman}
  \& {Armillotta}}{{Di Teodoro} et~al.}{2020}]{DiTeodoro2020}
{Di Teodoro} E.~M.,  {McClure-Griffiths} N.~M.,  {Lockman} F.~J.,
  {Armillotta} L.,  2020, \mn@doi [\nat] {10.1038/s41586-020-2595-z}, \href
  {https://ui.adsabs.harvard.edu/abs/2020Natur.584..364D} {584, 364}

\bibitem[\protect\citeauthoryear{{Dickey} \& {Lockman}}{{Dickey} \&
  {Lockman}}{1990}]{Dickey1990}
{Dickey} J.~M.,  {Lockman} F.~J.,  1990, \mn@doi [\araa]
  {10.1146/annurev.aa.28.090190.001243}, \href
  {https://ui.adsabs.harvard.edu/abs/1990ARA&A..28..215D} {28, 215}

\bibitem[\protect\citeauthoryear{{Dobler}, {Finkbeiner}, {Cholis}, {Slatyer}
  \& {Weiner}}{{Dobler} et~al.}{2010}]{Dobler2010}
{Dobler} G.,  {Finkbeiner} D.~P.,  {Cholis} I.,  {Slatyer} T.,   {Weiner} N.,
  2010, \mn@doi [\apj] {10.1088/0004-637X/717/2/825}, \href
  {https://ui.adsabs.harvard.edu/abs/2010ApJ...717..825D} {717, 825}

\bibitem[\protect\citeauthoryear{{Espada} et~al.,}{{Espada}
  et~al.}{2010}]{Espada2010}
{Espada} D.,  et~al., 2010, \mn@doi [\apj] {10.1088/0004-637X/720/1/666}, \href
  {https://ui.adsabs.harvard.edu/abs/2010ApJ...720..666E} {720, 666}

\bibitem[\protect\citeauthoryear{{Fielding} \& {Bryan}}{{Fielding} \&
  {Bryan}}{2022}]{Fielding2022}
{Fielding} D.~B.,  {Bryan} G.~L.,  2022, \mn@doi [\apj]
  {10.3847/1538-4357/ac2f41}, \href
  {https://ui.adsabs.harvard.edu/abs/2022ApJ...924...82F} {924, 82}

\bibitem[\protect\citeauthoryear{{GRAVITY Collaboration} et~al.,}{{GRAVITY
  Collaboration} et~al.}{2019}]{Gravity2019}
{GRAVITY Collaboration} et~al., 2019, \mn@doi [\aap]
  {10.1051/0004-6361/201935656}, \href
  {https://ui.adsabs.harvard.edu/abs/2019A&A...625L..10G} {625, L10}

\bibitem[\protect\citeauthoryear{{Gillmon}, {Shull}, {Tumlinson}  \&
  {Danforth}}{{Gillmon} et~al.}{2006}]{Gillmon2006}
{Gillmon} K.,  {Shull} J.~M.,  {Tumlinson} J.,   {Danforth} C.,  2006, \mn@doi
  [\apj] {10.1086/498053}, \href
  {https://ui.adsabs.harvard.edu/abs/2006ApJ...636..891G} {636, 891}

\bibitem[\protect\citeauthoryear{{Ginsburg} et~al.,}{{Ginsburg}
  et~al.}{2015}]{Ginsburg2015}
{Ginsburg} A.,  et~al., 2015, in {Iono} D.,  {Tatematsu} K.,  {Wootten} A.,
  {Testi} L.,  eds,  Astronomical Society of the Pacific Conference Series Vol.
  499, Revolution in Astronomy with ALMA: The Third Year. pp 363--364

\bibitem[\protect\citeauthoryear{{Greve}}{{Greve}}{2004}]{Greve2004}
{Greve} A.,  2004, \mn@doi [\aap] {10.1051/0004-6361:20031709}, \href
  {https://ui.adsabs.harvard.edu/abs/2004A&A...416...67G} {416, 67}

\bibitem[\protect\citeauthoryear{{Huang}, {Katz}, {Scannapieco}, {Cottle},
  {Dav{\'e}}, {Weinberg}, {Peeples}  \& {Br{\"u}ggen}}{{Huang}
  et~al.}{2020}]{Huang2020}
{Huang} S.,  {Katz} N.,  {Scannapieco} E.,  {Cottle} J.,  {Dav{\'e}} R.,
  {Weinberg} D.~H.,  {Peeples} M.~S.,   {Br{\"u}ggen} M.,  2020, \mn@doi
  [\mnras] {10.1093/mnras/staa1978}, \href
  {https://ui.adsabs.harvard.edu/abs/2020MNRAS.497.2586H} {497, 2586}

\bibitem[\protect\citeauthoryear{{Jonas} \& {MeerKAT Team}}{{Jonas} \& {MeerKAT
  Team}}{2016}]{Jonas2016}
{Jonas} J.,  {MeerKAT Team} 2016, in MeerKAT Science: On the Pathway to the
  SKA. p.~1, \mn@doi{10.22323/1.277.0001}

\bibitem[\protect\citeauthoryear{{Kanjilal}, {Dutta}  \& {Sharma}}{{Kanjilal}
  et~al.}{2021}]{Kanjilal2021}
{Kanjilal} V.,  {Dutta} A.,   {Sharma} P.,  2021, \mn@doi [\mnras]
  {10.1093/mnras/staa3610}, \href
  {https://ui.adsabs.harvard.edu/abs/2021MNRAS.501.1143K} {501, 1143}

\bibitem[\protect\citeauthoryear{{Koyama}, {Awaki}, {Kunieda}, {Takano}  \&
  {Tawara}}{{Koyama} et~al.}{1989}]{Koyama1989}
{Koyama} K.,  {Awaki} H.,  {Kunieda} H.,  {Takano} S.,   {Tawara} Y.,  1989,
  \mn@doi [\nat] {10.1038/339603a0}, \href
  {https://ui.adsabs.harvard.edu/abs/1989Natur.339..603K} {339, 603}

\bibitem[\protect\citeauthoryear{{Krumholz}}{{Krumholz}}{2014}]{Krumholz2014}
{Krumholz} M.~R.,  2014, \mn@doi [\mnras] {10.1093/mnras/stt2000}, \href
  {https://ui.adsabs.harvard.edu/abs/2014MNRAS.437.1662K} {437, 1662}

\bibitem[\protect\citeauthoryear{{Krumholz}, {McKee}  \&
  {Tumlinson}}{{Krumholz} et~al.}{2008}]{Krumholz2008}
{Krumholz} M.~R.,  {McKee} C.~F.,   {Tumlinson} J.,  2008, \mn@doi [\apj]
  {10.1086/592490}, \href
  {https://ui.adsabs.harvard.edu/abs/2008ApJ...689..865K} {689, 865}

\bibitem[\protect\citeauthoryear{{Krumholz}, {McKee}  \&
  {Tumlinson}}{{Krumholz} et~al.}{2009}]{Krumholz2009}
{Krumholz} M.~R.,  {McKee} C.~F.,   {Tumlinson} J.,  2009, \mn@doi [\apj]
  {10.1088/0004-637X/693/1/216}, \href
  {https://ui.adsabs.harvard.edu/abs/2009ApJ...693..216K} {693, 216}

\bibitem[\protect\citeauthoryear{{Lee}, {Stanimirovi{\'c}}, {Murray}, {Heiles}
  \& {Miller}}{{Lee} et~al.}{2015}]{Lee2015}
{Lee} M.-Y.,  {Stanimirovi{\'c}} S.,  {Murray} C.~E.,  {Heiles} C.,   {Miller}
  J.,  2015, \mn@doi [\apj] {10.1088/0004-637X/809/1/56}, \href
  {https://ui.adsabs.harvard.edu/abs/2015ApJ...809...56L} {809, 56}

\bibitem[\protect\citeauthoryear{{Leroy} et~al.,}{{Leroy}
  et~al.}{2015}]{Leroy2015}
{Leroy} A.~K.,  et~al., 2015, \mn@doi [\apj] {10.1088/0004-637X/814/2/83},
  \href {https://ui.adsabs.harvard.edu/abs/2015ApJ...814...83L} {814, 83}

\bibitem[\protect\citeauthoryear{{Liszt} \& {Gerin}}{{Liszt} \&
  {Gerin}}{2023}]{Liszt2023}
{Liszt} H.,  {Gerin} M.,  2023, \mn@doi [\apj] {10.3847/1538-4357/acae83},
  \href {https://ui.adsabs.harvard.edu/abs/2023ApJ...943..172L} {943, 172}

\bibitem[\protect\citeauthoryear{{Lockman}}{{Lockman}}{1984}]{Lockman1984}
{Lockman} F.~J.,  1984, \mn@doi [\apj] {10.1086/162277}, \href
  {https://ui.adsabs.harvard.edu/abs/1984ApJ...283...90L} {283, 90}

\bibitem[\protect\citeauthoryear{{Lockman} \& {McClure-Griffiths}}{{Lockman} \&
  {McClure-Griffiths}}{2016}]{Lockman2016}
{Lockman} F.~J.,  {McClure-Griffiths} N.~M.,  2016, \mn@doi [\apj]
  {10.3847/0004-637X/826/2/215}, \href
  {https://ui.adsabs.harvard.edu/abs/2016ApJ...826..215L} {826, 215}

\bibitem[\protect\citeauthoryear{{Lockman}, {Di Teodoro}  \&
  {McClure-Griffiths}}{{Lockman} et~al.}{2020}]{Lockman2020}
{Lockman} F.~J.,  {Di Teodoro} E.~M.,   {McClure-Griffiths} N.~M.,  2020,
  \mn@doi [\apj] {10.3847/1538-4357/ab55d8}, \href
  {https://ui.adsabs.harvard.edu/abs/2020ApJ...888...51L} {888, 51}

\bibitem[\protect\citeauthoryear{{Maccagni} et~al.,}{{Maccagni}
  et~al.}{2021}]{MacCagni2021}
{Maccagni} F.~M.,  et~al., 2021, \mn@doi [\aap] {10.1051/0004-6361/202141143},
  \href {https://ui.adsabs.harvard.edu/abs/2021A&A...656A..45M} {656, A45}

\bibitem[\protect\citeauthoryear{{Martini}, {Leroy}, {Mangum}, {Bolatto},
  {Keating}, {Sandstrom}  \& {Walter}}{{Martini} et~al.}{2018}]{Martini2018}
{Martini} P.,  {Leroy} A.~K.,  {Mangum} J.~G.,  {Bolatto} A.,  {Keating} K.~M.,
   {Sandstrom} K.,   {Walter} F.,  2018, \mn@doi [\apj]
  {10.3847/1538-4357/aab08e}, \href
  {https://ui.adsabs.harvard.edu/abs/2018ApJ...856...61M} {856, 61}

\bibitem[\protect\citeauthoryear{{McClure-Griffiths}, {Green}, {Hill},
  {Lockman}, {Dickey}, {Gaensler}  \& {Green}}{{McClure-Griffiths}
  et~al.}{2013}]{McClure-Griffiths2013}
{McClure-Griffiths} N.~M.,  {Green} J.~A.,  {Hill} A.~S.,  {Lockman} F.~J.,
  {Dickey} J.~M.,  {Gaensler} B.~M.,   {Green} A.~J.,  2013, \mn@doi [\apjl]
  {10.1088/2041-8205/770/1/L4}, \href
  {https://ui.adsabs.harvard.edu/abs/2013ApJ...770L...4M} {770, L4}

\bibitem[\protect\citeauthoryear{{McCourt}, {O'Leary}, {Madigan}  \&
  {Quataert}}{{McCourt} et~al.}{2015}]{McCourt2015}
{McCourt} M.,  {O'Leary} R.~M.,  {Madigan} A.-M.,   {Quataert} E.,  2015,
  \mn@doi [\mnras] {10.1093/mnras/stv355}, \href
  {https://ui.adsabs.harvard.edu/abs/2015MNRAS.449....2M} {449, 2}

\bibitem[\protect\citeauthoryear{{McKee} \& {Krumholz}}{{McKee} \&
  {Krumholz}}{2010}]{McKee2010}
{McKee} C.~F.,  {Krumholz} M.~R.,  2010, \mn@doi [\apj]
  {10.1088/0004-637X/709/1/308}, \href
  {https://ui.adsabs.harvard.edu/abs/2010ApJ...709..308M} {709, 308}

\bibitem[\protect\citeauthoryear{{McMullin}, {Waters}, {Schiebel}, {Young}  \&
  {Golap}}{{McMullin} et~al.}{2007}]{McMullin2007}
{McMullin} J.~P.,  {Waters} B.,  {Schiebel} D.,  {Young} W.,   {Golap} K.,
  2007, in {Shaw} R.~A.,  {Hill} F.,   {Bell} D.~J.,  eds,  Astronomical
  Society of the Pacific Conference Series Vol. 376, Astronomical Data Analysis
  Software and Systems XVI. p.~127

\bibitem[\protect\citeauthoryear{{Morganti}, {Oosterloo}  \&
  {Tsvetanov}}{{Morganti} et~al.}{1998}]{Morganti1998}
{Morganti} R.,  {Oosterloo} T.,   {Tsvetanov} Z.,  1998, \mn@doi [\aj]
  {10.1086/300236}, \href
  {https://ui.adsabs.harvard.edu/abs/1998AJ....115..915M} {115, 915}

\bibitem[\protect\citeauthoryear{{Saito} et~al.,}{{Saito}
  et~al.}{2022a}]{Saito2022}
{Saito} T.,  et~al., 2022a, \mn@doi [\apjl] {10.3847/2041-8213/ac59ae}, \href
  {https://ui.adsabs.harvard.edu/abs/2022ApJ...927L..32S} {927, L32}

\bibitem[\protect\citeauthoryear{{Saito} et~al.,}{{Saito}
  et~al.}{2022b}]{Saito2022b}
{Saito} T.,  et~al., 2022b, \mn@doi [\apj] {10.3847/1538-4357/ac80ff}, \href
  {https://ui.adsabs.harvard.edu/abs/2022ApJ...935..155S} {935, 155}

\bibitem[\protect\citeauthoryear{{Schneider} \& {Robertson}}{{Schneider} \&
  {Robertson}}{2017}]{Schneider2017}
{Schneider} E.~E.,  {Robertson} B.~E.,  2017, \mn@doi [\apj]
  {10.3847/1538-4357/834/2/144}, \href
  {https://ui.adsabs.harvard.edu/abs/2017ApJ...834..144S} {834, 144}

\bibitem[\protect\citeauthoryear{{Schneider}, {Robertson}  \&
  {Thompson}}{{Schneider} et~al.}{2018}]{Schneider2018}
{Schneider} E.~E.,  {Robertson} B.~E.,   {Thompson} T.~A.,  2018, \mn@doi
  [\apj] {10.3847/1538-4357/aacce1}, \href
  {https://ui.adsabs.harvard.edu/abs/2018ApJ...862...56S} {862, 56}

\bibitem[\protect\citeauthoryear{{Schneider}, {Ostriker}, {Robertson}  \&
  {Thompson}}{{Schneider} et~al.}{2020}]{Schneider2020}
{Schneider} E.~E.,  {Ostriker} E.~C.,  {Robertson} B.~E.,   {Thompson} T.~A.,
  2020, \mn@doi [\apj] {10.3847/1538-4357/ab8ae8}, \href
  {https://ui.adsabs.harvard.edu/abs/2020ApJ...895...43S} {895, 43}

\bibitem[\protect\citeauthoryear{{Schruba}, {Bialy}  \& {Sternberg}}{{Schruba}
  et~al.}{2018}]{Schruba2018}
{Schruba} A.,  {Bialy} S.,   {Sternberg} A.,  2018, \mn@doi [\apj]
  {10.3847/1538-4357/aac6c5}, \href
  {https://ui.adsabs.harvard.edu/abs/2018ApJ...862..110S} {862, 110}

\bibitem[\protect\citeauthoryear{{Seaquist} \& {Clark}}{{Seaquist} \&
  {Clark}}{2001}]{Seaquist2001}
{Seaquist} E.~R.,  {Clark} J.,  2001, \mn@doi [\apj] {10.1086/320448}, \href
  {https://ui.adsabs.harvard.edu/abs/2001ApJ...552..133S} {552, 133}

\bibitem[\protect\citeauthoryear{{Simonson}}{{Simonson}}{1973}]{Simonson1973}
{Simonson} S.~C. I.,  1973, \aap, \href
  {https://ui.adsabs.harvard.edu/abs/1973A&A....23...19S} {23, 19}

\bibitem[\protect\citeauthoryear{{Snowden} et~al.,}{{Snowden}
  et~al.}{1997}]{Snowden1997}
{Snowden} S.~L.,  et~al., 1997, \mn@doi [\apj] {10.1086/304399}, \href
  {https://ui.adsabs.harvard.edu/abs/1997ApJ...485..125S} {485, 125}

\bibitem[\protect\citeauthoryear{{Sofue} \& {Handa}}{{Sofue} \&
  {Handa}}{1984}]{Sofue1984}
{Sofue} Y.,  {Handa} T.,  1984, \mn@doi [\nat] {10.1038/310568a0}, \href
  {https://ui.adsabs.harvard.edu/abs/1984Natur.310..568S} {310, 568}

\bibitem[\protect\citeauthoryear{{Stanimirovi{\'c}}, {Murray}, {Lee}, {Heiles}
  \& {Miller}}{{Stanimirovi{\'c}} et~al.}{2014}]{Stanimirovic2014}
{Stanimirovi{\'c}} S.,  {Murray} C.~E.,  {Lee} M.-Y.,  {Heiles} C.,   {Miller}
  J.,  2014, \mn@doi [\apj] {10.1088/0004-637X/793/2/132}, \href
  {https://ui.adsabs.harvard.edu/abs/2014ApJ...793..132S} {793, 132}

\bibitem[\protect\citeauthoryear{{Su}, {Slatyer}  \& {Finkbeiner}}{{Su}
  et~al.}{2010}]{Su2010}
{Su} M.,  {Slatyer} T.~R.,   {Finkbeiner} D.~P.,  2010, \mn@doi [\apj]
  {10.1088/0004-637X/724/2/1044}, \href
  {https://ui.adsabs.harvard.edu/abs/2010ApJ...724.1044S} {724, 1044}

\bibitem[\protect\citeauthoryear{{Thompson}, {Quataert}, {Zhang}  \&
  {Weinberg}}{{Thompson} et~al.}{2016}]{Thompson2020}
{Thompson} T.~A.,  {Quataert} E.,  {Zhang} D.,   {Weinberg} D.~H.,  2016,
  \mn@doi [\mnras] {10.1093/mnras/stv2428}, \href
  {https://ui.adsabs.harvard.edu/abs/2016MNRAS.455.1830T} {455, 1830}

\bibitem[\protect\citeauthoryear{{Veilleux}, {Maiolino}, {Bolatto}  \&
  {Aalto}}{{Veilleux} et~al.}{2020}]{Veilleux2020}
{Veilleux} S.,  {Maiolino} R.,  {Bolatto} A.~D.,   {Aalto} S.,  2020, \mn@doi
  [\aapr] {10.1007/s00159-019-0121-9}, \href
  {https://ui.adsabs.harvard.edu/abs/2020A&ARv..28....2V} {28, 2}

\bibitem[\protect\citeauthoryear{{Wang} et~al.,}{{Wang}
  et~al.}{2021}]{Wang2021}
{Wang} G. C.~P.,  et~al., 2021, \mn@doi [\mnras] {10.1093/mnras/stab2800},
  \href {https://ui.adsabs.harvard.edu/abs/2021MNRAS.508.3754W} {508, 3754}

\bibitem[\protect\citeauthoryear{{Wannier}, {Lichten}  \& {Morris}}{{Wannier}
  et~al.}{1983}]{Wannier1983}
{Wannier} P.~G.,  {Lichten} S.~M.,   {Morris} M.,  1983, \mn@doi [\apj]
  {10.1086/160995}, \href
  {https://ui.adsabs.harvard.edu/abs/1983ApJ...268..727W} {268, 727}

\bibitem[\protect\citeauthoryear{{Werk} et~al.,}{{Werk}
  et~al.}{2014}]{Werk2014}
{Werk} J.~K.,  et~al., 2014, \mn@doi [\apj] {10.1088/0004-637X/792/1/8}, \href
  {https://ui.adsabs.harvard.edu/abs/2014ApJ...792....8W} {792, 8}

\bibitem[\protect\citeauthoryear{{Wong} et~al.,}{{Wong}
  et~al.}{2013}]{Wong2013}
{Wong} T.,  et~al., 2013, \mn@doi [\apjl] {10.1088/2041-8205/777/1/L4}, \href
  {https://ui.adsabs.harvard.edu/abs/2013ApJ...777L...4W} {777, L4}

\bibitem[\protect\citeauthoryear{{Yang}, {Ruszkowski}  \& {Zweibel}}{{Yang}
  et~al.}{2022}]{Yang2022}
{Yang} H. Y.~K.,  {Ruszkowski} M.,   {Zweibel} E.~G.,  2022, \mn@doi [Nature
  Astronomy] {10.1038/s41550-022-01618-x}, \href
  {https://ui.adsabs.harvard.edu/abs/2022NatAs...6..584Y} {6, 584}

\bibitem[\protect\citeauthoryear{Yuan, Krumholz  \& Martin}{Yuan
  et~al.}{2022}]{Yuan2023}
Yuan Y.,  Krumholz M.~R.,   Martin C.~L.,  2022, \mn@doi [\mnras]
  {10.1093/mnras/stac3241}, 518, 4084

\makeatother
\end{thebibliography}



\appendix

\section{Comparison of MeerKAT data with Green Bank Telescope}
\label{App:compari}
Here we compare the MeerKAT data for C1 and C2 to those obtained with the GBT as presented in \cite{DiTeodoro2018}. To conduct the comparison, we first re-bin all data sets to the same spatial and spectral bins and only consider data within the same velocity width (143-176 \kms for C1 and 242-319 \kms for C2) and spatial area ($(l,b)=(-1.76^{\circ}$ to $-1.96^{\circ}, -3.73^{\circ}$ to $-3.93^{\circ})$ for C1 and $(l,b)=(-2.36^{\circ}$ to $-2.46^{\circ}, 5.49^{\circ}$ to $5.59^{\circ})$) for C2. We then smooth the MeerKAT resolution to that of the GBT (569$''$) and compare the mass contained within the above-described areas. Our MeerKAT mass results are within 1$\%$ of the GBT mass results for both clouds. The differences in \hi{} mass reported in \autoref{tab:sourceinfo} compared to those from \cite{DiTeodoro2020} are due to the undefined nature of the clouds, with neither possessing distinct edges. As such, mass estimates depend on the spectral and spatial area over which the clouds' masses are calculated. The fact that the masses are consistent when these calculations are performed over the same PPV volume indicates that, despite the lack of short spacings, there is no significant missing flux in the MeerKAT data.

\section{Distance estimates}
\label{App:distance}
In this section, we consider how distance measurements affect mass estimates. To calculate the mass of a cloud, we require a distance measurement as described in \autoref{eqn:masshi} and \autoref{eqn:massh2}. The results presented in \autoref{tab:sourceinfo} use a distance of $8.2$ kpc (the distance from the Sun to the GC; \citealt{Gravity2019}) for all clouds. However, \citet{DiTeodoro2020} determine distances from the GC for each cloud using a kinematic model of the MW GC wind. Using these distances, we calculate the masses of each cloud as shown in \autoref{tab:appdistmass}.
\begin{table}
	\centering
	\caption{Derived masses of observed clouds. Distance is the cloud's distance from the Sun measured in kpc as determined by a kinematic wind model as described in \citet{DiTeodoro2020}. Atomic mass, $M_\mathrm{at}$, is the derived mass as traced by the \hi{} gas in solar masses from \autoref{eqn:masshi} using the distances in the Distance column. Molecular mass, $M_\mathrm{mol}$, is the derived mass as traced by the CO in solar masses from \autoref{eqn:massh2} using the distances in the Distance column.
 }
	\label{tab:appdistmass}
	\begin{tabular}{lcccr} 
		\hline
		Cloud ID & Distance & $M_\mathrm{at}$ & $M_\mathrm{mol}$\\ 
                & (kpc) & (M$_\odot$) & (M$_\odot$)\\
		\hline
		C1 & 8.7 & 385 & 443\\
		C2 &  9.7 & 1009 & 799 \\
            C3 & 8.3 & 3426 & -\\
		\hline
\end{tabular}
\end{table}


\bsp	
\label{lastpage}
\end{document}